\documentclass[aps,prb,epsf,twocolumn,showpacs,longbibliography]{revtex4-1}

\usepackage{amsmath,amssymb,graphics,epsfig,epstopdf,color,verbatim,ulem,braket,tabularx}
\usepackage[colorlinks,linkcolor=blue,citecolor=blue,urlcolor=blue]{hyperref}

\begin{document}

\title{Randomness induced spin-liquid-like phase in the spin-$1/2$ $J_1 - J_2$ triangular Heisenberg model}
\author{Han-Qing Wu$^1$}
\author{Shou-Shu Gong$^2$}
\email{shoushu.gong@buaa.edu.cn}
\author{D. N. Sheng$^1$}
\email{donna.sheng1@csun.edu}
\affiliation{
$^1$Department of Physics and Astronomy, California State University, Northridge 91330, USA\\
$^2$Department of Physics, Beihang University, Beijing, 100191, China
}

\begin{abstract}
We study the effects of bond randomness in the spin-$1/2$ $J_1-J_2$ triangular Heisenberg model
using exact diagonalization and density matrix renormalization group.
With increasing bond randomness, we identify a randomness induced spin-liquid-like phase without
any magnetic order, dimer order, spin glass order, or valence-bond glass order.
The finite-size scaling of gaps suggests the gapless nature of both spin triplet and singlet excitations,
which is further supported by the broad continuum of the dynamical spin structure factor.
By studying the bipartite entanglement spectrum of the system on cylinder geometry, we identify
the features of the low-lying entanglement spectrum in the spin-liquid-like phase, which may
distinguish this randomness induced spin-liquid-like phase and the intrinsic spin-liquid phase
in the clean $J_1 - J_2$ triangular Heisenberg model. We further discuss the nature of this
spin-liquid-like phase and the indication of our results for understanding spin-liquid-like
materials with triangular-lattice structure.
\end{abstract}


\date{\today}
\maketitle

\section{Introduction}

Frustrated quantum magnets realize a surprisingly rich place to explore the interplay between
classical orders and quantum fluctuations, which may lead to novel quantum phases and
unconventional quantum phase transitions~\cite{lacroix2013}.
One of the exotic quantum states is quantum spin liquid (QSL)~\cite{balents2010,savary2016,Norman2016,ZYi2017},
which breaks no spin rotational or lattice translational symmetry even at zero temperature and
exhibits fractionalized quasiparticles~\cite{wen1991,read1991} with  the emergent long-range
entanglement~\cite{chen2010}. QSL is now actively sought in quantum antiferromagnets with
frustrated and/or competing interactions~\cite{savary2016,Norman2016}, which may enhance quantum
fluctuations and suppress the ordering of magnetic moments. In experiment, many spin-$1/2$
antiferromagnetic materials on the frustrated lattices do not show any magnetic order down
to very low temperature; spin-liquid-like behaviors have also been observed in the neutron
scattering, NMR, and thermal conductance measurements (see Refs.~\onlinecite{savary2016},
\onlinecite{Norman2016}, \onlinecite{ZYi2017} and references therein).
Theoretical studies have indeed identified QSL states in particular parameter regime for some
microscopic models (see review articles Refs.~\onlinecite{savary2016}, \onlinecite{Norman2016}, \onlinecite{ZYi2017}).
However, it remains unclear whether these theoretical observed quantum states  explain the widely
reported spin-liquid-like behaviors in materials.

In reality, materials inevitably have defects and/or random disorder. For example, in the
triangular organic salt materials such as $\kappa$-(ET)$_2$Cu$_2$(CN)$_3$ and
EtMe$_3$Sb[Pd(dmit)$_2$]$_2$~\cite{shimizu2003, kurosaki2005, yamashita2008, minoru2008,yamashita2010},
the randomness of the spin degrees of freedom has been suggested as a consequence of the random
freezing of the electric-polarization degrees of freedom at low temperature~\cite{TSakai2014}.
In the kagome material herbertsmithite, the random substitution of magnetic Cu$^{2+}$ for
nonmagnetic Zn$^{2+}$ on the adjacent triangular layer would lead to the random modification
of the exchange couplings connecting the Cu$^{2+}$ on the kagome layer~\cite{freedman2010}.
The randomness may enhance quantum fluctuations and thus suppresses magnetic order. Very recently,
it has been proposed that the disorder even can generate long-range entanglement and
thus transform a classical non-Kramers spin ice into a QSL~\cite{savary2017}. The interplay
among frustration, quantum fluctuations, and randomness remains a largely open question in
the study of frustrated quantum magnetism, leaving the origin of the spin-liquid-like behaviors
in materials an intriguing question.

The pioneer corner stone of our understanding on randomness in quantum system is the
random singlet phase in the one-dimensional (1D) Heisenberg spin model, which represents
the infinite-randomness fixed-point (IRFP) in the strong-disorder renormalization group (SDRG)
and is universal for a broad class of spin chains~\cite{ma1979, dasgupta1980, fisher1992,
fisher1994}. The schematic picture of the random singlet state consists of pairs of spins which are
coupled together into singlets, where the long-range singlet bonds are much weaker than
the short ones and the singlet bonds cannot cross~\cite{ma1979, bhatt1982}. Later, extended 1D
chains and ladder systems with randomness have also been studied~\cite{melin2002,refael2002,
refael2004,quito2015,shu2016,shu2017}, in which other random phases such as the quantum Griffiths
phase~\cite{griffiths1969} and the spin-glass phase~\cite{binder1986} have been discovered.

In two dimensions (2D), Imry and Ma gave an argument for weak randomness, which suggests
that the ordered state is unstable against an arbitrarily small random field that is
directly coupled to the order parameter~\cite{imry1975}. In the strong-randomness case,
the IRFP has been found in a quantum Ising model~\cite{lesik2000,kovacs2011}, disordered contact
process~\cite{vojta2009pre}, or dissipative systems~\cite{vojta2009prb}. For the general
2D Heisenberg models, frustration is an intriguing ingredient that may lead to novel quantum
states. For example, while the N\'eel antiferromagnetic order persists up to the maximal
randomness in the bipartite square and honeycomb Heisenberg models without
frustration~\cite{laf2006, Kawamura2017}, the numerical SDRG calculation shows a large
spin formation in the frustrated Heisenberg models, suggesting a spin glass fixed
point~\cite{lin2003}. The potential effects of randomness in spin-liquid-like materials
have stimulated the exact diagonalization (ED) study on the frustrated triangular,
kagome, and honeycomb Heisenberg models~\cite{TSakai2014,kawamura2014,Kawamura2015,Kawamura2017},
in which the disordered phases displaying no magnetic or spin-glass order have been found
in the strong bond-randomness regime. The dynamical correlation and thermodynamic properties
of the random phases could be consistent with the gapless spin liquid scenario suggested
from experimental observations~\cite{TSakai2014,kawamura2014,Kawamura2015,Kawamura2017}.

Recently, a new triangular spin-liquid-like material YbMgGaO$_{4}$ has been reported~\cite{li2015tri,li2015,shen2016,paddison2017}.
The possible mixing of Mg$^{2+}$ and Ga$^{3+}$ ions in the material~\cite{li2015tri,li2015,li2017} has
stimulated further study on the randomness effects~\cite{luo2017,zhu2017,iaconis2017,kimchi2017,parker2018}.
More recently, another triangular-lattice compound YbZnGaO$_{4}$, which is a sister compound of YbMgGaO$_{4}$,
shows some spin-glass-like behaviors which may due to the disorder and frustration effects~\cite{ZhenMa2017}.
Since further-neighbor interaction in the material has been identified~\cite{paddison2017},
the nearest-neighbor model with disorder~\cite{TSakai2014,Kawamura2017} may not capture
the novel physics of such systems.
Inspired by the experimental indications, in this paper, considering the presence of  further-neighbor
couplings in materials, we study the bond randomness in the $J_{1}-J_{2}$ triangular Heisenberg model,
which would be more relevant to the randomness effects in the related materials.
In reality, spin-orbit coupling is strong in YbMgGaO$_4$ and YbZnGaO$_4$, which effectively induces anisotropic magnetic interactions. Nonetheless, theoretical studies have found that the microscopic model with only nearest-neighbor anisotropic interactions is always magnetically ordered~\cite{zhu2017, luo2017}. Competing interactions and disorder seem to be the dominant ingredients for the spin-liquid-like behavior~\cite{paddison2017,XinshuZang2018}. Thus here we study a simpler Heisenberg model with competing $J_2$ interaction and bond randomness so that we can use SU(2) symmetry  to deal with larger systems.
By using the ED and density matrix renormalization group (DMRG) calculation, we identify a randomness-induced spin-liquid-like (SLL) phase
that does not show any magnetic order, dimer order, spin-glass order, or valence-bond-glass (VBG) order.
The dynamical spin structure factor shows a broad continuum extending to the zero frequency, supporting
the gapless excitations obtained from the finite-size gap scaling. We also find the features of entanglement
spectrum in the SLL phase, which may distinguish the SLL phase and the intrinsic spin-liquid phase in the
$J_1 - J_2$ triangular Heisenberg model~\cite{Imada2014, bishop2015,SRWhite2015, DNSheng2015, iqbal2016, ian2016}.
The nature of this SLL phase appears to be consistent with the recently proposed 2D random singlet phase~\cite{kimchi2017}.
Finally, we discuss the relevance to the rare-earth triangular-lattice materials YbMgGaO$_4$ and YbZnGaO$_4$.

\begin{figure}[htp]
  \centering
  \includegraphics[width=0.45\textwidth]{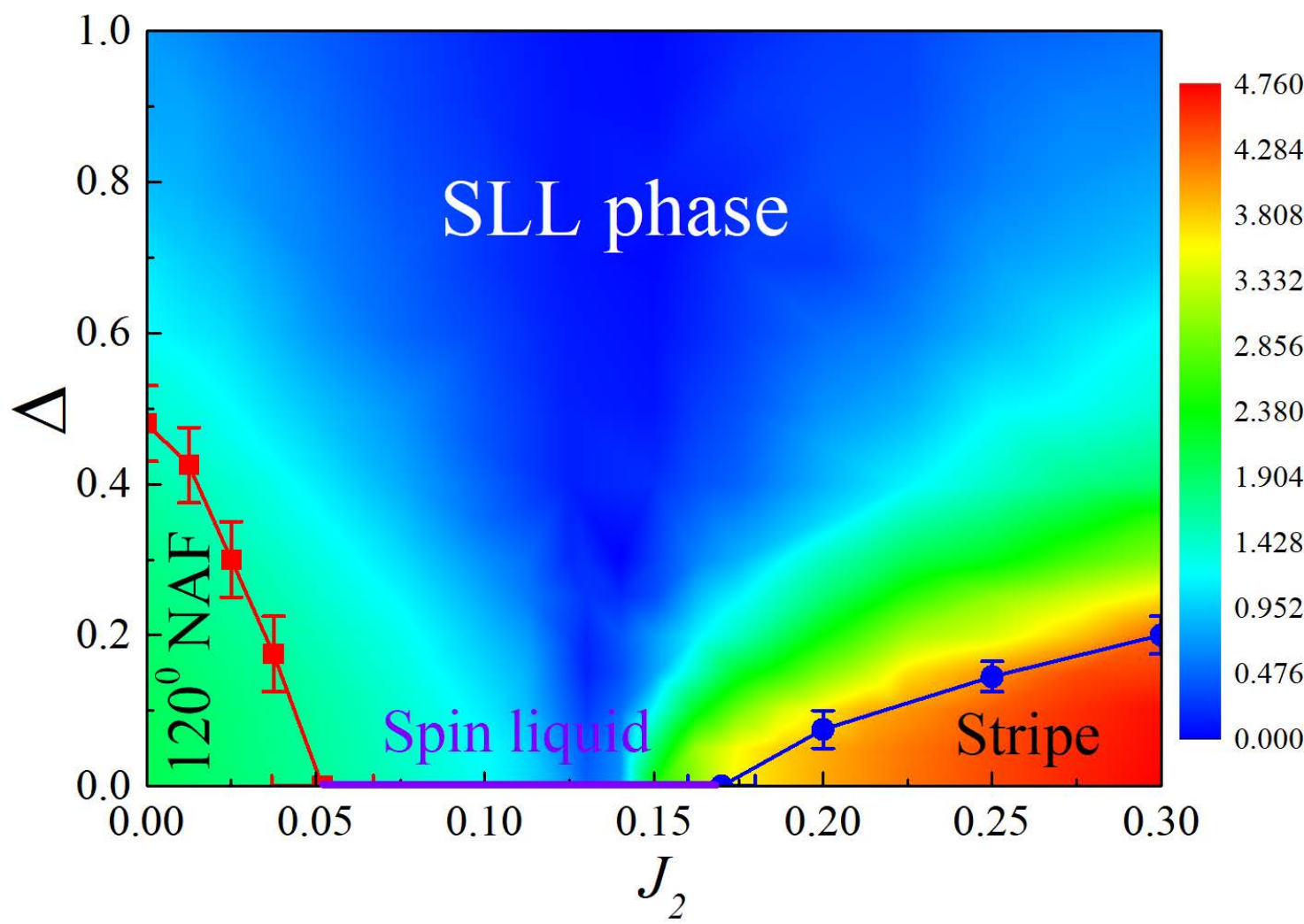}
  \caption{Contour plot of $|S(K) - S(M)|$ in the parameter space $J_{2}-\Delta$, where $S(K)$ and $S(M)$ are the spin structure factor for the $120^0$ N\'eel order and the stripe magnetic order obtained on the 24-site torus system. The definition of finite-size momentum points is shown in Appendix~\ref{App:FSClusters}. The solid points with error bars denote the phase boundaries between magnetic ordered and disordered phases, where the error bars are from the linear size-scaling of magnetic order parameters shown in Fig.~\ref{fig:MagneticOrders}. The possible quantum phase transition between the spin liquid phase and the randomness induced spin-liquid-like (SLL) phase is discussed in Sec.~\ref{Sec:SLandSLL}.}
  \label{fig:PhaseDiagram}
\end{figure}

\section{Model Hamiltonian and methods}

The Hamiltonian of the spin-$1/2$ $J_{1}-J_{2}$ Heisenberg model on the triangular lattice
with bond randomness reads
\begin{equation}
  \hat{H}=\sum_{\langle ij\rangle}J_{1}(1+\Delta\cdot\alpha_{ij})\hat{\mathbf{S}}_{i}
  \hat{\mathbf{S}}_{j} + \sum_{\langle\langle ij\rangle\rangle}J_{2}(1+\Delta\cdot\beta_{ij})
  \hat{\mathbf{S}}_{i}\hat{\mathbf{S}}_{j},
  \label{Eq:J1J2model}
\end{equation}
where $\alpha_{ij}$ and $\beta_{ij}$ are bond random variables which are uniformly distributed
in the interval $[-1,1]$, and $\Delta$ is the parameter to control the random interval
$[J_{i}(1-\Delta), J_{i}(1+\Delta)]$ of exchange interactions on each bond, $i=1, 2$ for the nearest
neighbor and the next-nearest neighbor. We use $\Delta\in[0,1]$ to ensure the antiferromagnetic coupling.
Here, we set $J_{1}=1$ as the energy constant.

We use ED and SU(2) DMRG~\cite{white1992,su2} to study this model. The finite-size clusters
we used are shown in Appendix~\ref{App:FSClusters}. To measure the possible orders in the system,
we define the high symmetry points in the first Brillouin zone (BZ), including the
$\Gamma$ point with ${\bf q} = (0,0)$, the $K$ point with ${\bf q} = (2\pi/3, 2\pi/\sqrt{3})$,
and the $M$ point with ${\bf q} = (\pi, \pi/\sqrt{3})$. While the 120$^0$ N\'eel order exhibits
the spin structure factor peak at the $K$ point, the stripe order has the peak at the $M$ point.
In the randomness case, we use 2000 (for smaller system sizes) to 20 (for the largest system size
with the number of lattice sites  $N=48$) in ED or DMRG torus calculation, and 15 independent samples
for YC6-24 and YC8-24 cylinders in DMRG calculation. We keep 2000 SU(2) states for torus
and 1200 SU(2) states for cylinder geometry in these  calculations. The truncation error
is less than $5\times 10^{-5}$. In the following, we use ``$\langle \rangle$" and ``[ ]" to
represent quantum mechanical expectation value and stochastic averaging, respectively.

\section{Numerical results}

\subsection{Melting the magnetic orders}

\begin{figure*}[!htp]
  \centering
  \includegraphics[width=0.85\textwidth]{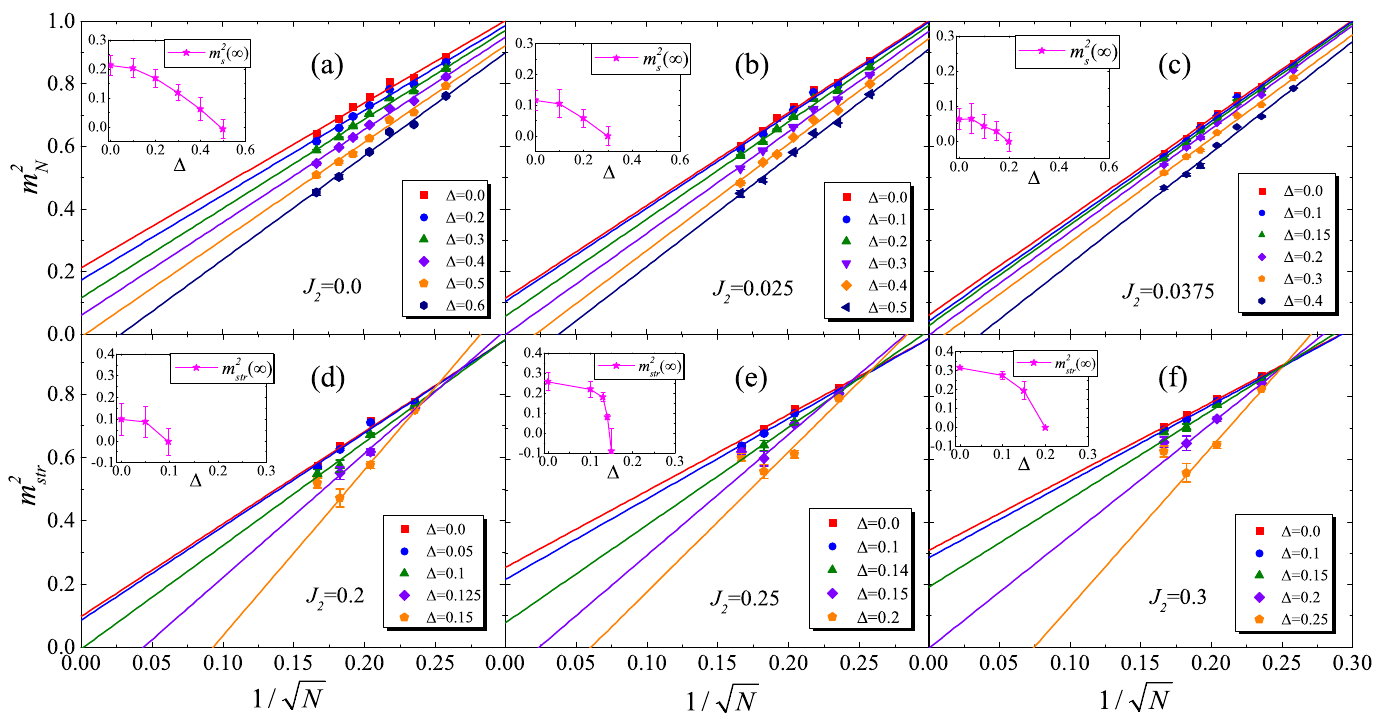}
  \caption{(Color online) Linear extrapolation of the square magnetization of (a-c) the
  $120^0$ N\'eel order and (d-f) the stripe order versus $1/\sqrt{N}$ ($N$ is the total
  site number). The insets show the extrapolated order parameters as a functions of bond
  randomness strength $\Delta$. The vanishing orders with bond randomness can be used to
  estimate the phase boundaries between the magnetic order phases and the nonmagnetic phase.
  In the stripe phase, the $m^2_{str}$ of the $36$-site torus shows some deviation from
  other system sizes due to the finite-size geometry effects (see Appendix~\ref{App:FSClusters}).
  }
  \label{fig:MagneticOrders}
\end{figure*}

In the absence of randomness, the spin-$1/2$ $J_1 - J_2$ triangular Heisenberg model exhibits
an intermediate spin-liquid phase for $0.07 \lesssim J_{2} \lesssim 0.15$ according to the previous
study~\cite{Imada2014, bishop2015, SRWhite2015, DNSheng2015, iqbal2016, ian2016},
which is sandwiched between the $120^{0}$ N\'eel phase and the stripe phase.
First of all, we identify the intermediate nonmagnetic phase from the
vanishing magnetic orders that are extrapolated to the thermodynamic limit using the
torus data up to $36$ sites (see Appendix~\ref{App:J1J2Model}).
Our estimations qualitatively agree with the previous results although the ED data slightly
overestimate the intermediate regime because of the finite-size effects.
Next, we focus on the system with bond randomness.

In the magnetic order phases, bond randomness is not directly coupled with the order
parameter and it has been found that only a finite bond randomness may kill the magnetic order
~\cite{TSakai2014,laf2006,Kawamura2017}. In order to quantitatively characterize how the ordering strength
decreases with bond randomness, we introduce two magnetic order parameters. The first is the square sublattice magnetization for the $120^{0}$ N\'eel antiferromagnetic (NAF) phase~\cite{TSakai2014, Kawamura2015}
\begin{equation}
  m_{N}^{2}=\frac{1}{3}\sum_{\alpha=1}^{3}\left[\frac{1}{(N/6)(N/6+1)}\left\langle \left(\sum_{i\in\alpha}\hat{\mathbf{S}}_{i}\right)^2\right\rangle\right],
  \label{eq:msqrt120AF}
\end{equation}
where $\alpha=1,2,3$ represents the three sublattices of the $120^{0}$ order (which is labeled
by the three different colors in Appendix~\ref{App:FSClusters}).
For the classical $120^{0}$ N\'eel state, the spins in the same sublattice order
ferromagnetically and the spins in the different sublattices are in the same
plane with $120^{0}$ angle structure. So actually we have normalized $m_{s}^{2}$
to $1$ in the classical case by using the expectation value $(N/6)(N/6+1)$
of the total spin operator in the sublattice. In the quantum case, the definition of
Eq.~\eqref{eq:msqrt120AF} describes the residual order after considering quantum fluctuations. The second is the square sublattice magnetization for the stripe antiferromagnetic phase~\cite{Kawamura2015}
\begin{equation}
  m_{str}^{2}=\frac{1}{2}\sum_{\beta=1}^{2}\left[\frac{1}{(N/4)(N/4+1)}\left\langle \left(\sum_{i\in\beta}\hat{\mathbf{S}}_{i}\right)^2\right\rangle\right],
  \label{eq:msqrtstripe}
\end{equation}
where $\beta=1,2$ represents the two sublattices of the stripe order. $m_{str}^{2}$ has also been
normalized to 1 in the classical stripe phase. According to the spin-wave theory~\cite{neuberger1989},
the magnetic orders follow the size scaling behavior
\begin{equation}
  m_{N/str}^{2}=m_{s/str}^{2}(\infty)+\frac{c_{1}}{\sqrt{N}}+\frac{c_{2}}{N}+\cdots.
\end{equation}
We use the leading behavior of this scaling function $1/\sqrt{N}$ to
estimate the magnetic order strength in the thermodynamic limit through finite-size scaling.

In Fig.~\ref{fig:MagneticOrders}, we show the linear extrapolation of the magnetic orders using torus geometry up to $36$ sites.
To consider the two competing magnetic orders simultaneously, we choose the cluster geometries that are compatible with both
the $120^0$ order and the stripe order.
For this reason, we choose only the 12-, 18-, 24-, and 30-site clusters for the size scaling of $m_{str}^2$ as shown in Figs.~\ref{fig:MagneticOrders}(d-f).
Both orders are suppressed by increasing randomness.
Up to some critical values, the bond randomness kills the magnetic orders.
The system undergoes a quantum phase transition to a randomness-induced nonmagnetic phase.
Then we can estimate the phase boundaries between the magnetic order phases and the nonmagnetic phase in the $J_{2}-\Delta$ phase diagram of Fig.~\ref{fig:PhaseDiagram}.

\subsection{Randomness induced spin-liquid-like phase}

\begin{figure}[htp]
  \centering
  \includegraphics[width=0.48\textwidth]{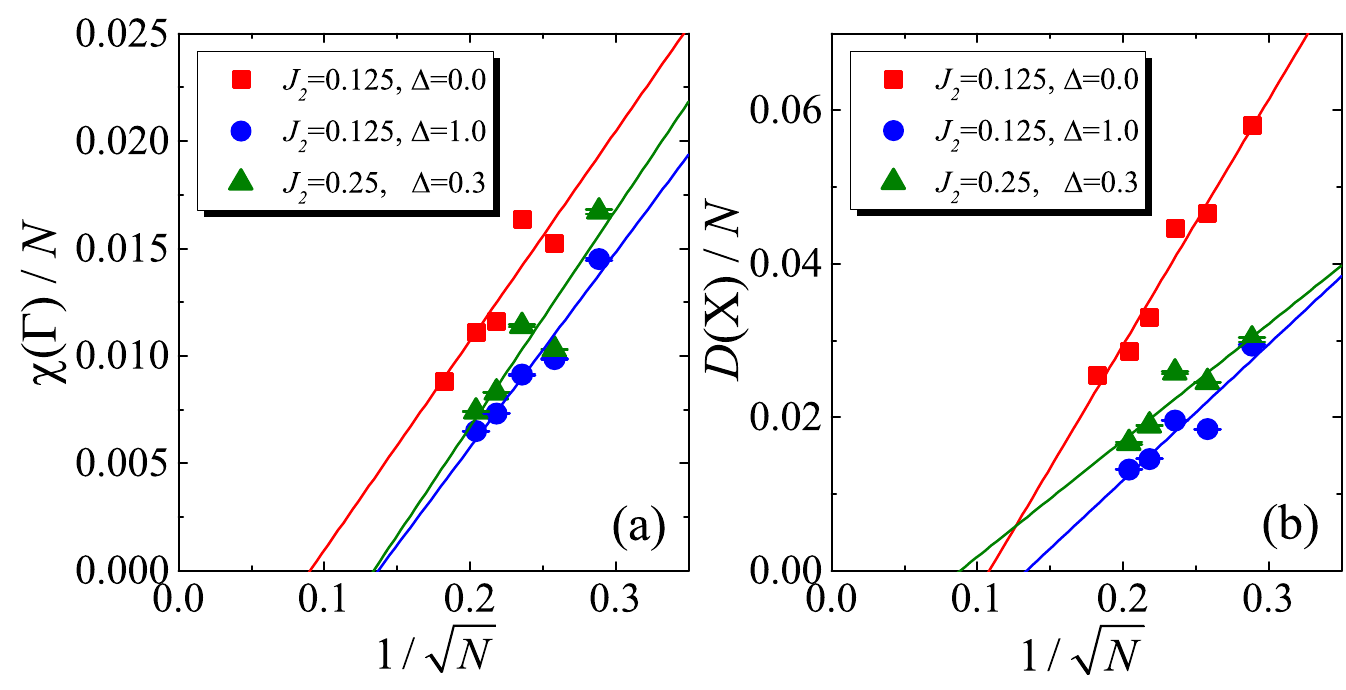}
  \caption{(Color online) Linear extrapolation of (a) chiral and (b) dimer orders vs system
size $1/\sqrt{N}$ in the nonmagnetic regime with or without bond randomness. Both orders go to zero in the
thermodynamic limit. Point X is the momentum point where the dimer structure factor shows the
maximum value, see Appendix~\ref{App:DimerCorr}.}
  \label{fig:ChiETDimET}
\end{figure}

In this section, we will focus on characterizing the SLL phase. We first show that
there is no long-range chiral or dimer order. For detecting the possible orders,
we define the structure factor for the scalar chiral correlation as
\begin{equation}
\begin{split}
  \chi(\mathbf{q})&=\frac{1}{N}\sum_{ij}e^{-i\mathbf{q}\mathbf{r}_{ij}}\left[\langle\hat{\chi}_{i}\hat{\chi}_{j}\rangle\right],\\
  \hat{\chi_{i}}&=\hat{\mathbf{S}}_{i}\cdot\left(\hat{\mathbf{S}}_{i+\mathbf{a}_{1}}\times\hat{\mathbf{S}}_{i+\mathbf{a}_{2}}\right),
\end{split}
\end{equation}
and the structure factor for the dimer correlation as
\begin{equation}
\begin{split}
  D(\mathbf{q})&=\frac{1}{3N}\sum_{ij}\sum_{pq}e^{-i\mathbf{q}\mathbf{r}_{ip,jq}}\left[\left\langle\hat{\mathbf{B}}_{ip}\hat{\mathbf{B}}_{jq}
  \right\rangle\right],\\
  \hat{\mathbf{B}}_{ip} &= \hat{\mathbf{S}}_{i}\hat{\mathbf{S}}_{i+p}-\langle\hat{\mathbf{S}}_{i}\hat{\mathbf{S}}_{i+p}\rangle,
  \label{eq:DimerBi}
\end{split}
\end{equation}
where $i+p$ means the nearest-neighbor site of $i$-site along $\mathbf{a}_{1}, \mathbf{a}_{2}, -\mathbf{a}_{1}+\mathbf{a}_{2}$
direction for $p=1,2,3$ respectively. $\mathbf{a}_{1}$ and $\mathbf{a}_{2}$ are the primitive vectors on the triangular lattice.
$\mathbf{r}_{ip,jq}$ means the displacement between centers of two bonds, see Appendix~\ref{App:DimerCorr}. In Fig.~\ref{fig:ChiETDimET},
we show the finite-size scaling of the peak value of the chiral and dimer structure factors. Apparently, as the bond randomness increases, these two structure factors become weaker, which does not show any ordering tendency both in the clean limit and the large randomness limit.

\begin{figure}[htp]
  \centering
  \includegraphics[width=0.48\textwidth]{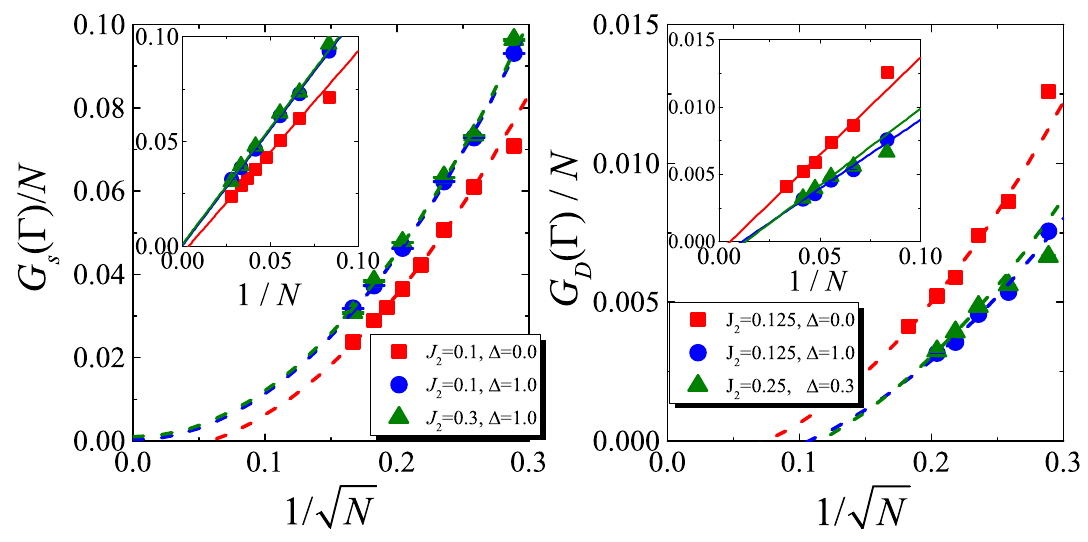}
  \caption{(Color online) Finite-size scaling of the spin glass and valence bond glass
  structure factor peak. The insets show the linear extrapolation of glass orders as a
  function of $1/N$. The dashed lines are guides to the eye using the fitting results in the insets.}
  \label{fig:ETGlass}
\end{figure}

\begin{figure}[htp]
  \centering
  \includegraphics[width=0.48\textwidth]{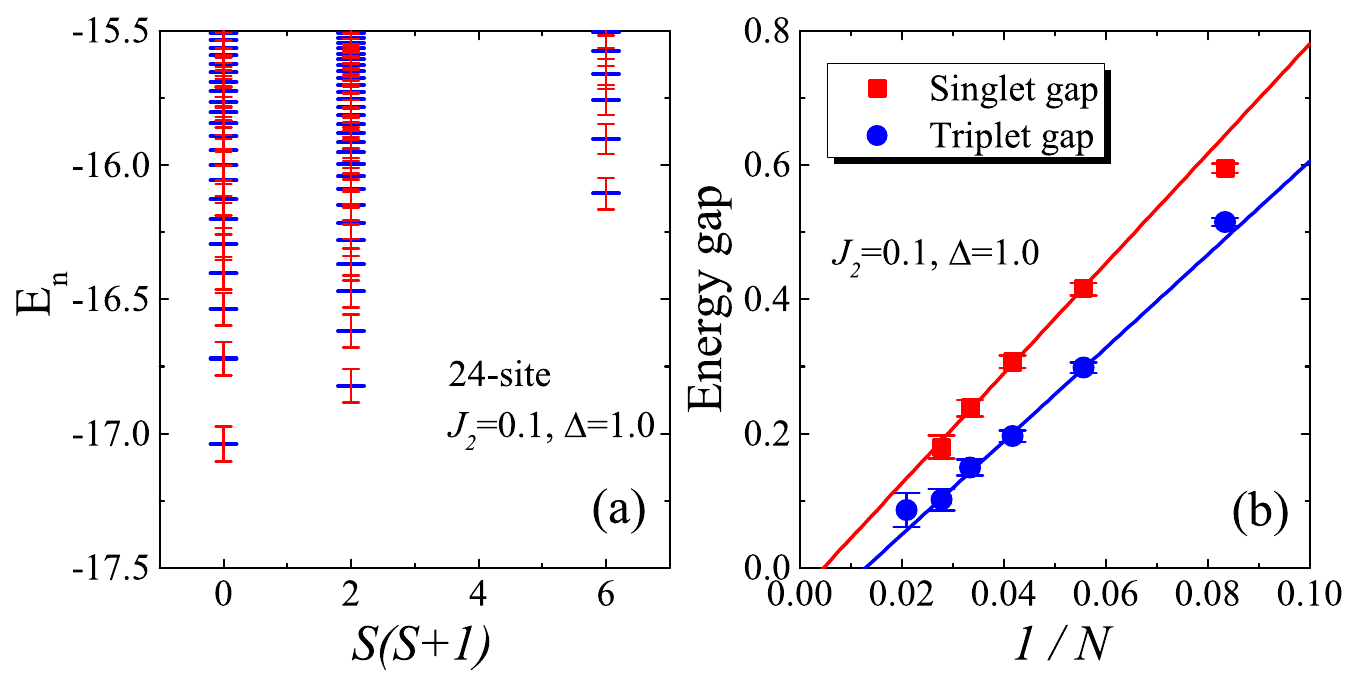}
  \caption{(Color online) (a) Energy spectrum for $J_{2}=0.1, \Delta=1.0$ on the $24$-site torus as a function of
  total spin $S(S+1)$. (b) Linear scaling of the singlet and triplet gap vs $1/N$ for $J_{2}=0.1, \Delta=1.0$.}
  \label{fig:EngyGap}
\end{figure}

In magnetic systems, randomness may induce glass orders at low temperature such as the spin glass~\cite{binder1986} and
valence bond glass~\cite{tarzia2008,singh2010}, which have short-range order but do not show long-range order. For example, the
spin glass state has the vanished total magnetization
$M = \frac{1}{N} \sum_{i} \left[ \langle \hat{\mathbf{S}}_i \rangle \right] = 0$
but the nonzero spin glass order
$\bar{q} = \frac{1}{N} \sum_{i} \left[ \langle \hat{\mathbf{S}}_i \rangle^2 \right] \neq 0$.
For detecting the possible glass order, we define the structure factor for the
square spin correlation
\begin{equation}
  G_{S}(\mathbf{q})=\frac{1}{N}\sum_{ij}e^{-i\mathbf{q}\mathbf{r}_{ij}}
  \left[\langle\hat{\mathbf{S}}_{i}\hat{\mathbf{S}}_{j}\rangle^2\right],
\end{equation}
which can be used to detect the spin glass order. In our calculation, we find the peak of
$G_{S}(\mathbf{q})$ at the $\Gamma$ point with $\mathbf{q} = (0, 0)$, which is the spin-glass
susceptibility and can be used as the spin-glass order parameter~\cite{nonomura1995, oitmaa2001}.
If the peak value increases with system size $N$ equal to or faster than a linear behavior,
the order could be finite in the thermodynamic limit. In our calculation, we find that
$G_{S}(\Gamma)/N$ appropriately scales to zero with both $1/\sqrt{N}$ and $1/N$,
as we can see in Fig.~\ref{fig:ETGlass}(a), indicating the vanished spin glass order.
In the 2D Ising spin glass phase, the spin glass order scales with
$\langle \bar{q}^2(L) \rangle - \langle \bar{q}^2(\infty) \rangle \propto L^{-1/2}$~\cite{thomas2011,rubin2017,xu2018},
which is quite different from this triangular model, where the order seems more natural to scale with $1/N$.
Although the spin glass order grows slightly with increased randomness on finite-size system, the order parameter
actually drops faster with increasing system size.
Clearly, for both $J_2=0.1, \Delta=1.0$ and $J_2=0.3, \Delta=1.0$ cases, the linearly extrapolated values are zero within numerical error.
The absence of the  spin glass order in the SLL phase has also been found in other frustrated
Heisenberg models with bond randomness~\cite{TSakai2014,kawamura2014,Kawamura2017}.

\begin{figure*}[!htp]
\centering
\includegraphics[width=0.85\textwidth]{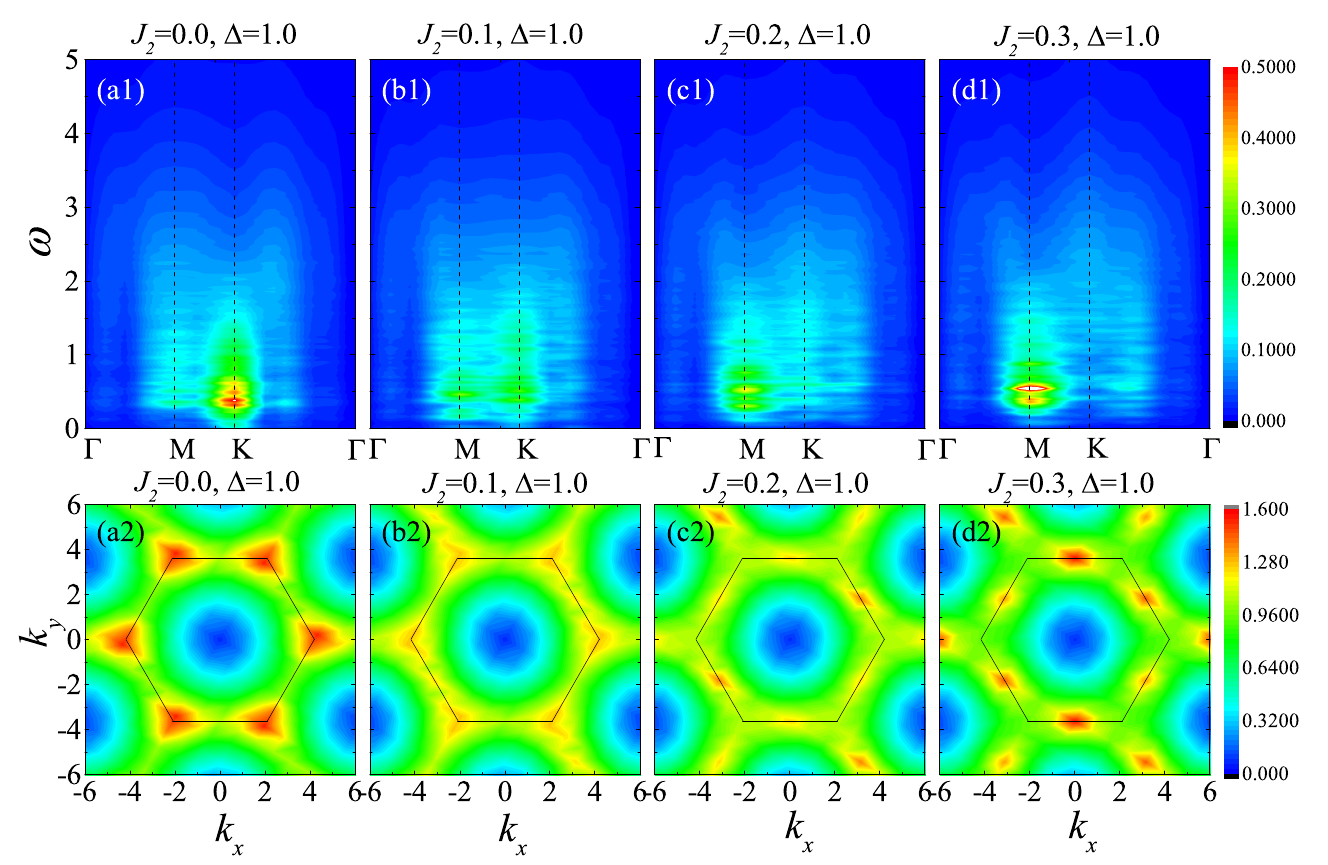}
\caption{(Color online) (a1)-(d1) Dynamical spin structure factor $S^{zz}({\bf q}, \omega)$
along the high-symmetry path $\Gamma\rightarrow M\rightarrow K\rightarrow\Gamma$ in the
momentum space. We show the results of only the $24$-site cluster here. The results on other
system sizes, such as 12 and 18, are similar to the $24$-site cluster. (a2)-(d2) Static spin structure factor
obtained on the YC8-24 cylinder using DMRG. We take the middle $8\times 8$ sites in the
cylinder to do the Fourier transform. As the cylinder geometry does not respect the
$C_{6}$ rotation symmetry, the three $M$ points are not equivalent.}
\label{fig:TDSpinCorr}
\end{figure*}

Similar to the spin glass order, we could define the structure factor for the VBG correlation as
\begin{equation}
G_{D}(\mathbf{q})=\frac{1}{3N}\sum_{ij}\sum_{pq}e^{-i\mathbf{q}\mathbf{r}_{ip,jq}}
\left[\langle\hat{\mathbf{B}}_{ip}\hat{\mathbf{B}}_{jq}\rangle^2\right],
\end{equation}
where $\hat{\mathbf{B}}_{ip}$ has been defined in Eq.~\eqref{eq:DimerBi}.
The VBG structure factor also shows the peak at the $\Gamma$ point.
Interestingly, the VBG peak at the $\Gamma$ point seems to decrease with
growing randomness as shown in Fig.~\ref{fig:ETGlass}(b), which indicates
the absent VBG order in the SLL phase.

For further characterization of the SLL phase, we study the energy spectrum and the excitation
gaps. In Fig.~\ref{fig:EngyGap}(a), we show a random averaged energy spectrum on the $24$-site
torus. The eigenvalues appear to be continuously distributed in the energy landscape.
In both the ED torus and the DMRG cylinder calculations, the random averaged ground state is the
nondegenerate spin singlet state (the ground state has probability to be in the $S = 1$ sector in some random
distributions) and the averaged first excited state is the spin triplet state.
In Fig.~\ref{fig:EngyGap}(b), we show that in the SLL phase both the singlet gap
$\Delta_{SS}=E_{1}(S=0)-E_{0}(S=0)$ and the triplet gap
$\Delta_{ST}=E_{0}(S=1)-E_{0}(S=0)$ drop fast and seem to go to vanishing, suggesting gapless excitations.

\begin{figure}[htp]
\includegraphics[width=0.48\textwidth]{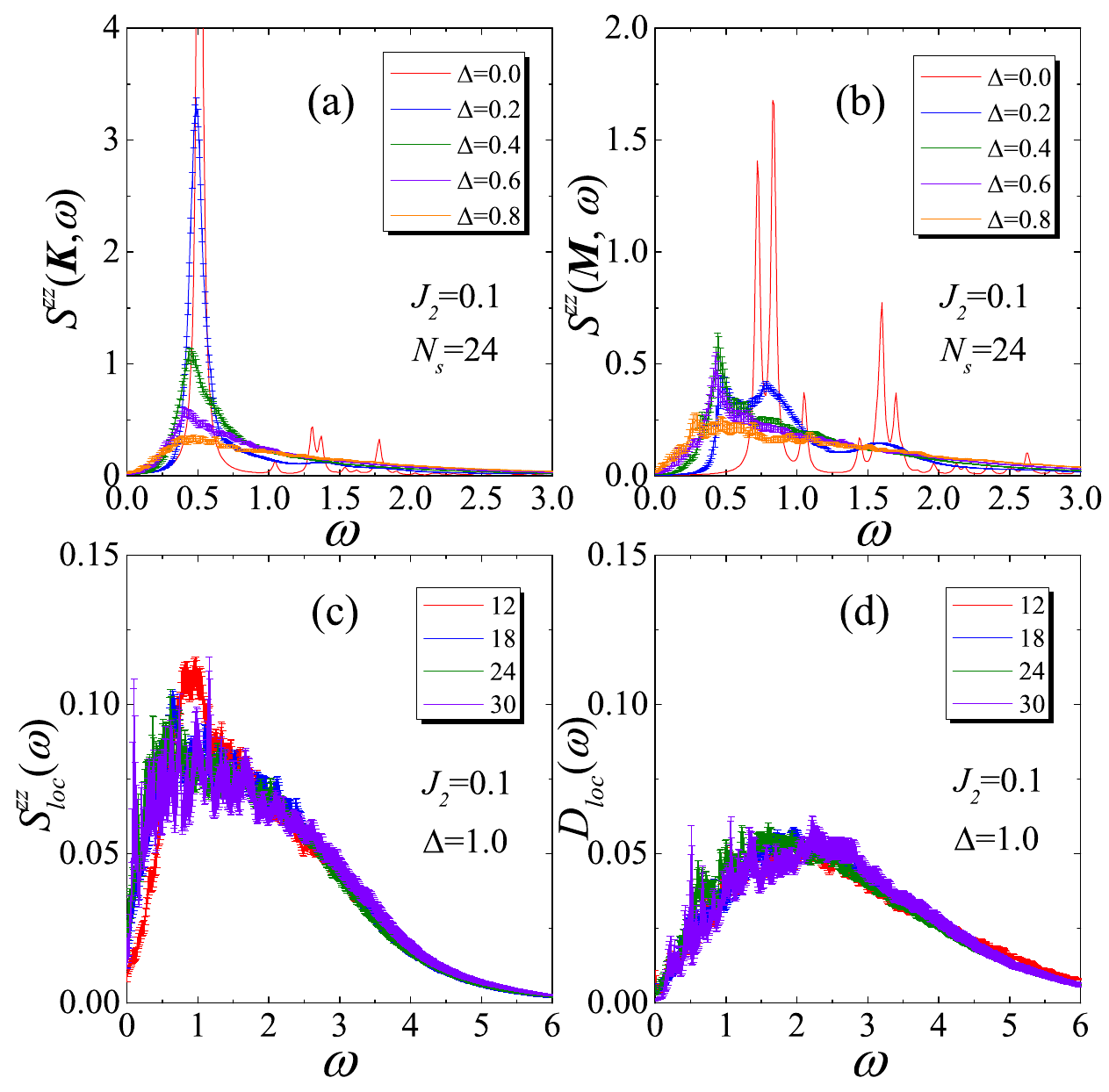}
\caption{(Color online) (a-b) The dynamical structure factor at the $K$
and $M$ points for $J_2 = 0.1$ on the $24$-site torus system with different
bond randomness strength. (c-d) The momentum-integrated dynamical spin and
dimer correlations for $J_2 = 0.1, \Delta = 1.0$ in the SLL phase on different system
sizes.}
\label{fig:TDLocCorr}
\end{figure}

Next, we study the dynamical spin correlation using ED simulation. We define the dynamical spin structure factor as
\begin{equation}
S^{zz}(\mathbf{q},\omega)=\sum_{n}\left[|\langle \psi_n | \hat{S}^z_{\bf q} | \psi_0 \rangle|^2 \delta(\omega-(E_n-E_0))\right],
\end{equation}
where $\hat{S}_{\mathbf{q}}^{z}=(1/N)\sum_{i}e^{-i\mathbf{q}\mathbf{r}_{i}}\hat{S}_{i}^{z}$
is the Fourier transform of the $z$-component of spin operator, $|\psi_n\rangle$ is the eigenstate
of the Hamiltonian with energy $E_n$, and $|\psi_0\rangle$ is the ground state with energy $E_0$.
The dynamical spin structure factor describes the correlations in both space and time,
which can be studied by inelastic neutron scattering (INS) or X-ray Raman scattering.
In the Lanczos iteration method~\cite{Lanczos1950,SaadBook2011}, the dynamical structure
factor can be computed by continued fraction expansion~\cite{gagliano1987} using Lanczos coefficients and rewritten as
\begin{equation}
\begin{split}
  &S^{zz}(\mathbf{q},\omega)\\
  &=-\frac{1}{\pi}\lim\limits_{\eta\rightarrow 0}\text{Im}\left[\bra{\psi_{0}}\left(\hat{S}_{\mathbf{q}}^{z}\right)^{\dagger}\frac{1}{\omega+E_{0}-\hat{H}+i\eta}\hat{S}_{\mathbf{q}}^{z}\ket{\psi_{0}}\right],\\
  &=-\frac{1}{\pi}\lim\limits_{\eta\rightarrow 0}\text{Im}\left[\cfrac{\bra{\psi_{0}}\left(\hat{S}_{\mathbf{q}}^{z}\right)^{\dagger}\hat{S}_{\mathbf{q}}^{z}\ket{\psi_{0}}}
  {z-a_{0}-\cfrac{b_{1}^{2}}{z-a_{1}-\cfrac{b_{2}^{2}}{z-a_{2}\cdots}}}\right],
  \label{Eq:SzzTD}
\end{split}
\end{equation}
where $z=\omega+E_{0}+i\eta$, $a_{i}$ and $b_{i+1}$ are the diagonal and sub-diagonal
elements of the tridiagonal Hamiltonian matrix obtained by the Lanczos method with initial
vector $\hat{S}_{\mathbf{q}}^{z}|\psi_0 \rangle$. The Lorentz broadening factor we use is $\eta=0.02$.

In Figs.~\ref{fig:TDSpinCorr}(a1)-\ref{fig:TDSpinCorr}(d1), we show the dynamical structure factor $S^{zz}({\bf q}, \omega)$ at different $J_{2}$
along the high-symmetry path $\Gamma\rightarrow M\rightarrow K\rightarrow\Gamma$ in the
large randomness case with $\Delta = 1.0$. For small $J_{2}$, we can see a broad maxima
at the $K$ point with a low frequency, showing the short-range spin correlation dominated
by the $120^0$ N\'eel type. With increasing $J_{2}$, the spectrum weight gradually
transfers to the $M$ point, which indicates the dominant stripe-like short-range correlation
for $J_2 \gtrsim 0.2$. This behavior can be seen more clearly from the static spin structure factor
$S(\mathbf{q})=(1/N)\sum_{ij}e^{\mathbf{q}\mathbf{r}_{ij}}[\langle\hat{S}_{i}\hat{S}_{j}\rangle]$
shown in Figs.~\ref{fig:TDSpinCorr}(a2)-\ref{fig:TDSpinCorr}(d2), where the broad peak at the $K$ point transfers
its weight to the $M$ point as $J_{2}$ increases. Even with strong bond randomness, frustration
seems to still affect short-range spin correlation. In the dynamical structure factor,
we also find that the broad finite spectrum extends to zero frequency, supporting the
gapless excitations suggested in Fig.~\ref{fig:EngyGap}(b).

For further insight into the $K$ point and $M$ point at the edge of the BZ, we show the
dramatic changing of the dynamic spectrum as a function of randomness strength $\Delta$
in Fig.~\ref{fig:TDLocCorr}(a) and \ref{fig:TDLocCorr}(b), starting from the $J_1 - J_2$ spin liquid regime.
In small randomness, we see a sharp peak at the $K$ point with the frequency
$\omega \sim 0.5$, which seems to signature coherently propagating magnon excitation.
Note that this sharp peak might be due to possible strong finite-size effects~\cite{
DNSheng2015,ian2016,gong2017} in the intermediate $J_1 - J_2$ spin liquid phase.
Meanwhile, the spectrum at the
$M$ point exhibits several weaker peaks. As the randomness increases, the peak at the
$K$ point transfers its weight to lower and higher frequencies, keeping a broad maxima
near $\omega \sim 0.5$. On the other hand, the peak at the $M$ point also becomes broad
but shifts to the lower frequency. When the randomness is sufficiently large, a broad
continuum spectrum with an exponentially decaying high-frequency tail not only appears at the $K$ and $M$ points
but also stretches to other wave vectors near the edge of the Wagner-Seitz Brillouin zone,
which is quite different from the magnon like excitations.

In order to consider the finite-size effects, we show the local or momentum integrated dynamical spin-spin correlation
with different system sizes in Fig.~\ref{fig:TDLocCorr} (c), which is defined as
\begin{equation}
\begin{split}
  &S_{loc}^{zz}(\omega)=S_{ii}^{zz}(\omega)=\int d\mathbf{q} S^{zz}(\mathbf{q},\omega)=\\
  &-\frac{1}{\pi}\lim\limits_{\eta\rightarrow 0}\text{Im}\left[\bra{\psi_{0}}\hat{S}_{i}^{z}\frac{1}{\omega+E_{0}-\hat{H}+i\eta}\hat{S}_{i}^{z}\ket{\psi_{0}}\right],
  \label{Eq:SzzLoc}
\end{split}
\end{equation}
where $i$ is the real-space lattice site. Although randomness breaks translation symmetry,
it can be approximately restored if the number of random samples is large enough and thus we can
take $i$ as any lattice site. We have also calculated the local dynamical dimer correlation
in Fig.~\ref{fig:TDLocCorr} (d), which is defined as
\begin{equation}
\begin{split}
  &D_{ii}(\omega)=\\
  &-\frac{1}{\pi}\lim\limits_{\eta\rightarrow 0}\text{Im}\left[\bra{\psi_{0}}\hat{\mathbf{B}}_{i}^{\dagger}\frac{1}{\omega+E_{0}-\hat{H}+i\eta}\hat{\mathbf{B}}_{i}\ket{\psi_{0}}\right],
\end{split}
\end{equation}
where $\hat{\mathbf{B}}_{i}$ is defined in Eq.~\eqref{eq:DimerBi}. The two
local dynamical correlations share the similar behaviors including
the broad spectrum and the finite density in the zero frequency.
Mostly significantly, the finite-size effects in the SLL phase are
not manifest even though we use small clusters due to the limit of
system size.

In the recent INS measurements on the triangular spin-liquid material
YbMgGaO$_4$~\cite{shen2016,paddison2017,li2017nc}, broad continuum spin excitations
have been reported. While the high-energy spin excitations between $0.25$ and $1.5$ meV
have been conjectured to be related with either a gapless spinon Fermi surface~\cite{shen2016}
or the nearest-neighbor resonating valence bond correlations~\cite{li2017nc}, the low-energy
excitations down to $0.02 $meV~\cite{li2017nc} seem to include crucial information on the
origin of the spin-liquid-like behaviors in the material, which is currently debated
between an intrinsic spin liquid and a disorder-induced mimicry of a spin
liquid~\cite{shen2016,paddison2017,luo2017,zhu2017,kimchi2017}. By considering the
scenario of the disorder-induced spin-liquid-like phase, we compare our numerical
results in the SLL phase with the INS data of YbMgGaO$_4$. The SLL phase shows some
similar behaviors of dynamical spin correlations with the experiment of YbMgGaO$_4$,
including the broadly spread spectral weights in the Brillouin zone and the suppressed
spectral intensities near the $\Gamma$ point~\cite{paddison2017}. In the INS intensity
data, the maxima at the $K$ point above $0.5$ meV shifts to the $M$ point below
$0.1$ meV~\cite{paddison2017,li2017nc}. The broad low-energy excitation maxima at the
$M$ point could be consistent with our SLL phase with a small $J_2$ coupling as shown
in Fig.~\ref{fig:TDSpinCorr}(d1).

\begin{figure}[htp]
  \centering
  \includegraphics[width=0.48\textwidth]{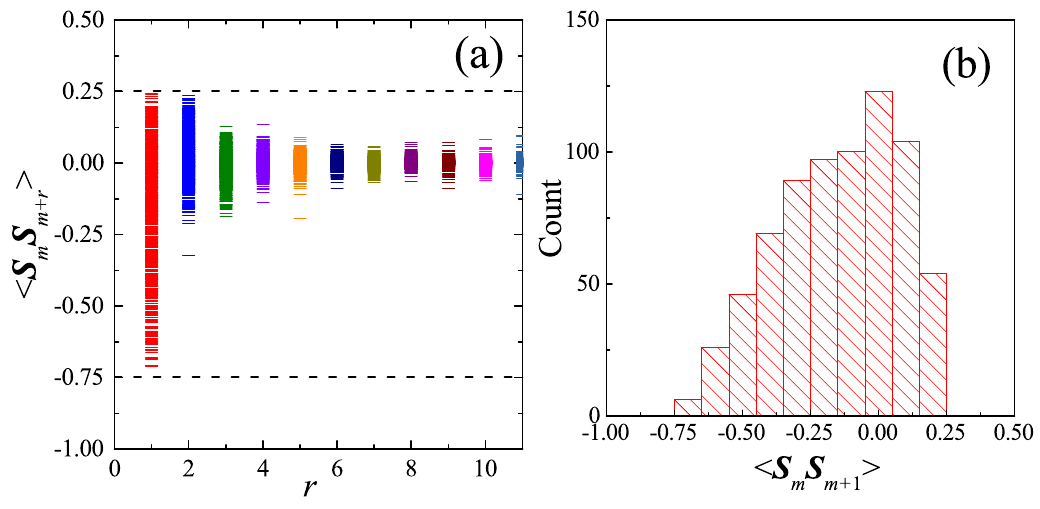}
  \caption{(Color online) (a) Spin-spin correlations along the $x$ direction on the YC6-24 cylinder.
  The reference site $m$ is taken in the middle of the cylinder. $r$ is the distance of the two sites
  along the $x$ direction. We show the results of $720$ independent random samples in the figure. The dashed
  lines show the lower and upper bounds of spin-spin correlation.
  (b) The histogram of nearest-neighbor spin-spin correlation obtained from $720$ independent random
  samples. We take $0.1$ as the bar unit of the $x$ axis. The $y$ axis denotes the count number that
  the random sample gives the spin correlation value in the range of the given unit bar. Here, the
  next-nearest-neighbor interaction and bond randomness strength are chosen as $J_{2}=0.125, \Delta=1.0$.}
  \label{fig:BondEngy}
\end{figure}

Therefore, we identify a gapless SLL phase in the presence of strong bond randomness.
In this SLL phase, we have not observed any conventional order or glass-type order.
For further understanding on this phase, we calculate the sample distribution of spin correlation
$\langle \mathbf{\hat{S}}_{i} \mathbf{\hat{S}}_{j} \rangle$ as shown in Fig.~\ref{fig:BondEngy} (a).
Interestingly, at the larger distance side $r\geq 6$, the width of correlation distribution saturates
to some finite value, which indicates the emergent long-range correlations between two spins with
near equal probability of both positive and negative signs for different randomness configurations.
To look into the details of nearest-neighbor correlation, we show the histogram of its distribution
in Fig.~\ref{fig:BondEngy} (b). Compared with 1D random singlet phase in the bond randomness Heisenberg
chain (see Appendix~\ref{App:Histogram}), this distribution in the SSL phase shows a low probability
near $-\frac{3}{4}J$. Different values of the next-nearest neighbor $J_2$ would not change this
behavior. The geometry frustration and the high coordination number $z=6$ in the triangular lattice
may play an important role here.

\subsection{$J_1-J_2$ spin liquid and the SLL phase}
\label{Sec:SLandSLL}

\begin{figure}[htp]
  \centering
  \includegraphics[width=0.48\textwidth]{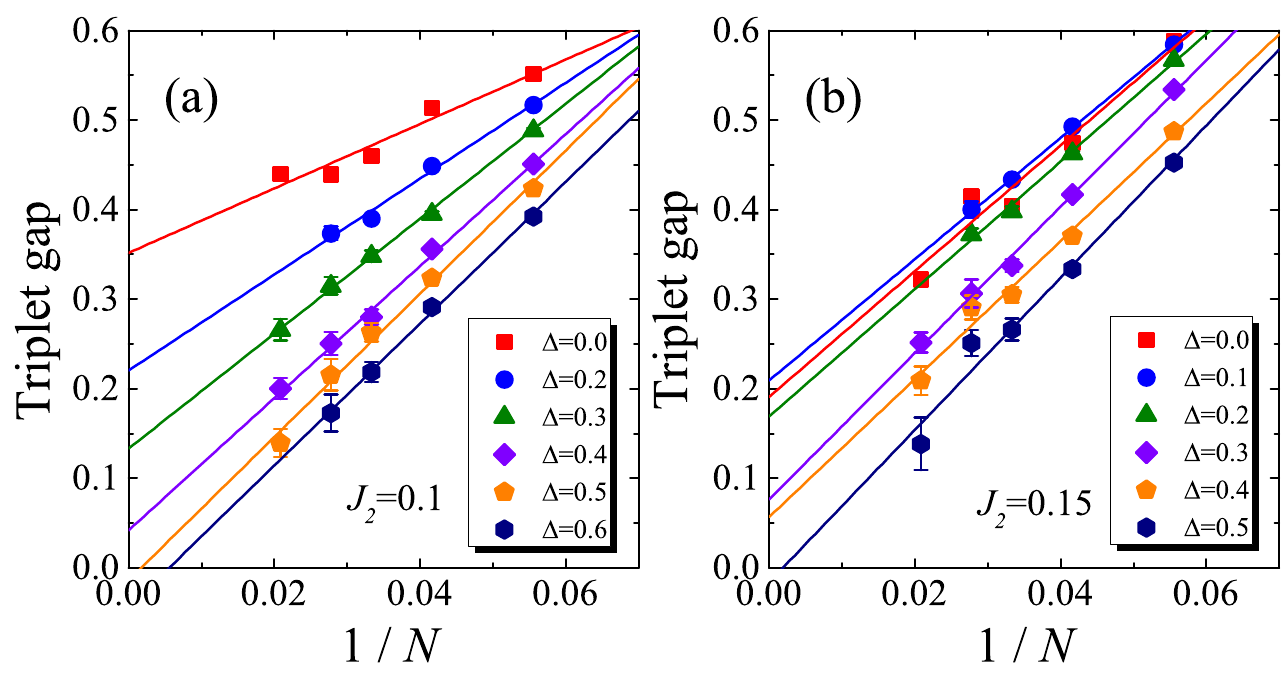}
  \caption{(Color online) Linear size scaling of the spin triplet gap with inverse system size $1/N$ at
  (a) $J_{2}=0.1$ and (b) $J_{2}=0.15$. We see the blend down behavior with growing randomness on finite-size system.}
  \label{fig:SpinGap}
\end{figure}

In this section, we study the difference between the $J_{1}-J_{2}$ spin liquid and
the SLL phase. In the absence of randomness, the nature of the $J_1-J_2$ spin liquid
is still debated between a gapless Dirac spin liquid and a gapped spin liquid~\cite{Imada2014,
bishop2015,SRWhite2015,DNSheng2015,iqbal2016,ian2016,gong2017}. We calculate the triplet gap on
the torus clusters up to 48 sites (see Appendix~\ref{App:J1J2Model}), nonetheless the small-size data
may not draw a conclusive evidence to show whether the gap is finite or not. If the
gap is finite, we may expect a quantum phase transition from the gapped QSL to the gapless SLL
phase, as suggested in Fig.~\ref{fig:SpinGap}. However, if the ED calculation suffers
from strong finite-size effects and the spin liquid turns out to be gapless~\cite{Imada2014,iqbal2016},
our present size scaling may not correctly show the phase transition.

\begin{figure}
\includegraphics[width=0.48\textwidth]{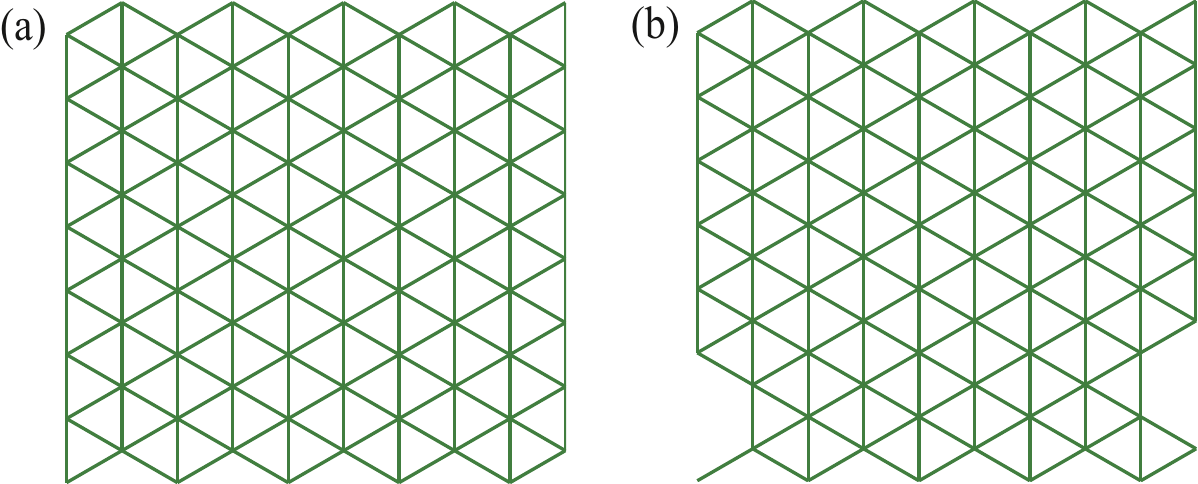}
\caption{(Color online) The YC8 cylinder with the even (a) and odd (b) boundary
conditions in the $x$ direction. In the odd boundary condition (b), a spin-$1/2$ site is removed
in each open edge.}
\label{fig:YC}
\end{figure}

\begin{figure}[htp]
  \centering
  \includegraphics[width=0.5\textwidth]{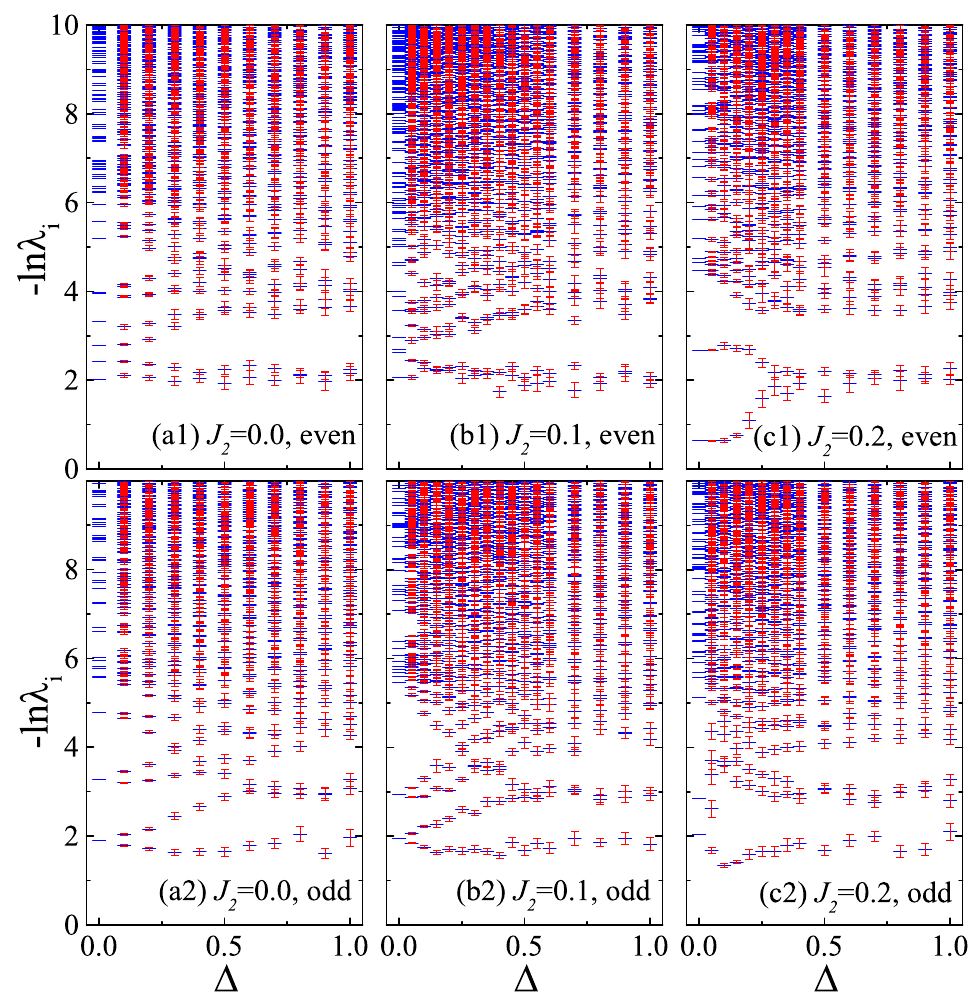}
  \caption{(Color online) Entanglement spectra in the (a1-c1) even and (a2-c2) odd boundary conditions
  obtained on the YC8-24 cylinder using DMRG. $\lambda_i$ are the eigenvalues of the reduced density
  matrix. The error bars are estimated from $15$ independent randomness samples.}
  \label{fig:YC8ES}
\end{figure}

Since the QSL and the SLL state may have different entanglement structure, we calculate
the entanglement spectrum on the cylinder geometry with two different open edges in the $x$
direction. We denote the even boundary as the usual boundary conditions shown in
Fig.~\ref{fig:YC}(a) and the odd boundary by removing a spin-$1/2$ site on each open edge of
the cylinder, as shown in Fig.~\ref{fig:YC}(b). In Fig.~\ref{fig:YC8ES}, we show the
entanglement spectra obtained on the YC8-24 cylinder. First of all, we analyze the spectrum
in the SLL phase. In the even boundary shown in Figs.~\ref{fig:YC8ES}(a1)-\ref{fig:YC8ES}(c1), the spectrum
always show a two-fold near degeneracy separated by a finite gap from the higher levels. The two-fold
eigenvalues are identified as coming from the $S=0$ and the $S=1$ sectors ($S$ is the good quantum
number of total spin for the subsystem). In the odd boundary shown in Figs.~\ref{fig:YC8ES}(a2)-\ref{fig:YC8ES}(c2),
one level with $S = 1/2$ and two levels with $S = 1/2, 3/2$ are found in the low-lying spectrum.
These features in both boundary conditions seem to be independent of $J_2$ for systems with
large strength of randomness, which might be used to characterize the SLL phase.

Next, we investigate the change of entanglement spectrum with randomness, starting from the
$J_1 - J_2$ spin liquid. Since the characterization of the spin liquid phase in the even boundary
conditions is likely to have large finite-size effects~\cite{DNSheng2015,gong2017}, here we
consider the spectrum in the odd boundary conditions as shown in Fig.~\ref{fig:YC8ES}(b2).
In the absence of randomness, the entanglement spectrum has a double degeneracy for all the
eigenvalues~\cite{SRWhite2015,DNSheng2015}. With increasing randomness, the two lowest eigenvalues
split. For large randomness, we can see one level with $S = 1/2$ and two levels with $S = 1/2, 3/2$,
which are separated from the higher spectrum. This feature for the SLL phase appears at $\Delta \sim 0.5$.
We have also checked the entanglement spectrum of the YC6-24 cylinder and got a similar result as the YC8-24.
In the kagome Heisenberg model, a possible phase transition induced by randomness between the clean
kagome spin liquid and the SLL phase has been suggested at $\Delta \sim 0.4$~\cite{Kawamura2015},
where the randomness sampling starts to have probability for the triplet ground state. In the
ED calculation of the triangular model with $J_2 = 0.1$, we find the probability for triplet ground
state at $J_2 \gtrsim 0.6$, which is close to $0.5$. The consistency between these different pictures
suggests that the entanglement spectrum may be used as a characterization to distinguish the spin liquid and the SLL phase.

\section{Summary and discussion}

By using the exact diagonalization (ED) and density matrix renormalization group (DMRG)
techniques, we have studied the spin-$1/2$ $J_{1}-J_{2}$ triangular Heisenberg model with
bond randomness in both $J_1$ and $J_2$ couplings. In the absence of the randomness,
the model has two magnetic order phases and a spin-liquid phase between them~\cite{Imada2014,
bishop2015,SRWhite2015,DNSheng2015,iqbal2016,ian2016}. This spin liquid phase may even extend
to the anisotropic model that could be relevant to materials~\cite{zhu2018}. By turning
on the bond randomness, we find a randomness-induced spin-liquid-like (SLL) phase above a
finite randomness strength $\Delta$ for a given $J_2$, as shown in the phase
diagram Fig.~\ref{fig:PhaseDiagram}. This SLL phase does not show any spin, dimer, spin-glass,
or valence-bond-glass order in our finite-size scaling. The spin triplet and singlet gaps also
seem to be vanishing after the finite-size scaling. These static properties suggest a gapless
spin-liquid-like phase induced by bond randomness, which is supported by the dynamical spin
structure factor $S^{zz}({\bf q}, \omega)$. In the SLL phase, $S^{zz}({\bf q}, \omega)$ shows
a broad continuum in both momentum and frequency space. With growing $J_2$, the broad maxima at
the $K$ point transfers its weight to the $M$ point, showing that frustration affects
short-range spin correlations even in the presence of strong randomness. We compare the dynamical
spin correlations of the SLL phase with the inelastic neutron scattering (INS) data of the
spin-liquid-like triangular material YbMgGaO$_4$. The dynamical spectrum of the SLL phase
with a small $J_2$ coupling could be consistent with the INS data of the low-energy excitations
of YbMgGaO$_4$, which shows the dominant broad maxima at the $M$ point~\cite{paddison2017,li2017nc}.

For studying randomness effects in the disordered $J_1 - J_2$ spin liquid, we examine the
bipartite entanglement spectrum on cylinder geometry. We find the low-lying spectrum features
in the SLL phase, which seems independent of $J_2$ and may characterize the random phase.
This feature of entanglement spectrum appears at $\Delta \simeq 0.5$, which may suggest a
phase transition from the spin liquid to the SLL phase and deserves more further studies.
Before further discussion, we would like to remark that although most of our calculations are based on
the ED method, we have pushed the system size as large as we can. Due to the limit of system size,
one should not interpret all the results as the final answer; however, we believe that our main
results are convincing, including the gapless nonmagnetic behavior of the SLL phase, the absent
glass-type orders, and the characteristic features of dynamical spin structure factor.
In the absence of $J_2$ coupling, the bond randomness has been studied in previous ED calculation, which also proposed a spin-liquid-like phase with growing randomness~\cite{TSakai2014}. Based on our phase diagram Fig.~\ref{fig:PhaseDiagram}, it seems that the disordered phase extends to a large region with finite $J_2$. No other disorder phase such as spin glass has been found.

Furthermore, we would like to discuss the nature of the SLL phase. In 2D systems, randomness
may induce different quantum phases, with some examples  such as a spin glass~\cite{binder1986},
VBG~\cite{tarzia2008, singh2010}, and quantum Griffiths phase~\cite{griffiths1969,yu2006}. These phases have
been found in the diluted and random-graph-like systems, which are quite different from
our model with bond coupling randomness and a perfect lattice geometry. For the SLL phase in
this $J_1 - J_2$ triangular model, our results suggest that spin glass and VBG phases are
unlikely. The numerical SDRG analysis for frustrated Heisenberg models suggested a
spin glass fixed point~\cite{lin2003}, which however seems inconsistent with our result and
recent numerical studies on other frustrated models~\cite{TSakai2014,kawamura2014,Kawamura2015,Kawamura2017}.
In a recent theoretical paper by Kimchi {\textit{et al.}}, the authors have studied the effects
of bond randomness on 2D valence bond solid and spin liquid states~\cite{kimchi2017}. They
found that the bond randomness inevitably leads to the nucleation of topological defects
with spin-$1/2$ when destructing the valence bond order, which would yield gapless spin
excitations and the short-ranged VBG order would be unstable. The SLL phase found in our
numerical calculation, which shows gapless spin excitations and vanished VBG order, appears
to be in agreement with the proposed state in Ref.~\onlinecite{kimchi2017}. The next check of this SLL phase could
be the thermodynamic properties such as specific heat and susceptibility, which we leave for future study.

Finally, we would like to make some remarks about the application of our results to experiments.
For YbMgGaO$_4$, bond randomness may not be weak~\cite{li2017}, and second-neighbor interaction may play an important role for
the observations of experiments~\cite{paddison2017}. Theoretical calculations found that the spin anisotropic interactions
may not drive a spin-liquid-like behavior but support magnetic ordering~\cite{luo2017, zhu2017}.
By considering a minimum model to study the effects of competing interaction and disorder, we find that
the dynamical structure factor of the spin-liquid-like phase with a small $J_2$ agrees with the INS data of YbMgGaO$_4$.
The gapless excitations and the absence of the spin glass order are also consistent with experimental observations.
All these results indicate a consistent  description of the spin-liquid-like phase on the ground state of YbMgGaO$_4$ from our minimum model.
In this $J_1 - J_2$ model, we do not find a spin glass order in the presence of bond randomness.
For understanding the spin-glass-like freezing in materials such as YbZnGaO$_4$~\cite{ZhenMa2017}, other spin anisotropic couplings  may play important roles, which deserves further study.

{\it Note added.} Recently, we became aware of an interesting work~\cite{Liu2018}, which studied a spin-$1/2$ $J-Q$ model on the square lattice with bond randomness using
 quantum Monte Carlo. The authors also found a disorder-induced spin-liquid-like phase, which was suggested as a random singlet phase.

\begin{acknowledgments}

D.N.S thanks Leon Balents for suggesting the problem.
We thank Wen-An Guo, Dao-Xin Yao, Rong-Qiang He, and Zi Yang Meng for fruitful discussions.
We also acknowledge extensive discussions with Itamar Kimchi.
H. Q. W would like to thank Wei Zhu for helpful discussions about block-diagonalization using symmetries.
H. Q. W also would like to thank the Magic-II platform at Shanghai Supercomputer Center.
This research is supported by the National Science Foundation through Grants
No. PREM DMR-1828019 (H.Q.W.) and  No. DMR-1408560 (D.N.S.).  S.S.G. is supported by the  National Natural Science Foundation of China through Grants No. 11834014 and No. 11874078,
and start-up funding support from Beihang University.

\end{acknowledgments}

\appendix

\section{Finite-size clusters}
\label{App:FSClusters}

\begin{figure}[htbp]
  \centering
  \includegraphics[width=0.45\textwidth]{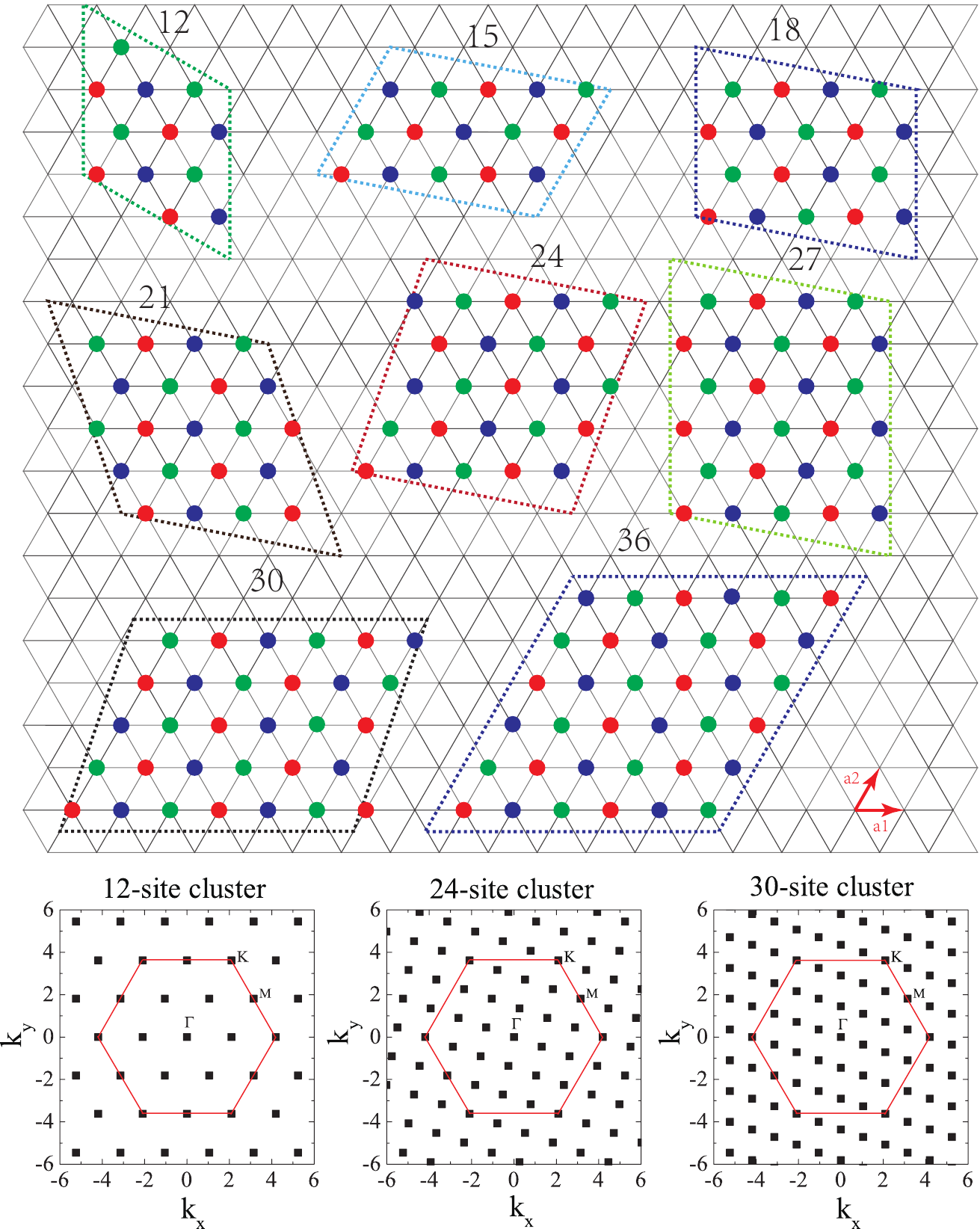}
  \caption{Most of the finite-size clusters used in the numerical calculations. The red, green and olive
  solid points represent three sublattices of $120^{0}$ AF order or $\sqrt{3}\times\sqrt{3}$ magnetic
  order. $\mathbf{a}_{1}=(a,0)$ and $\mathbf{a}_{2}=(a/2,\sqrt{3}a/2)$ are primitive vectors. Here we
  set the lattice constant or nearest-neighbor bond length $a=1$ as a unit of length. The dashed lines
  which connect the bond centers of the triangular lattice in the 24-site cluster form a kagome lattice.
  The bottom three figures show the finite-size points in momentum space. In addition, 18- and 48- site
  rhombic clusters can be easily obtained by expanding $6\times3$ and $8\times6$ primitive cells.
  The 18-site rhombic cluster was used in the calculation of singlet and triplet gaps in Figs.~\ref{fig:AppSinglet} and \ref{fig:AppTriplet}.}
  \label{fig:AppClusters}
\end{figure}

In this paper, we use both ED and DMRG to do the tori calculations. These tori are made of two
dimension clusters (which are shown in Fig.~\ref{fig:AppClusters}) under periodic boundary conditions.
In order to get unbiased extrapolations, the geometries of tori are important. Since $120^0$ Neel order and stripe order are the two competing magnetic phases,
they need to be considered on an equal footing. Therefore, almost all the
 geometries (except for the 48-site geometry) we chose are commensurate to the $120^{0}$ antiferromagnetic order,
i.e. they have two K momentum points in the Brillouin zone (BZ). And all the clusters with even sites
are also commensurate to the collinear or stripe order.
We also note that the   36-site and 48-site clusters have both three M  points in the BZ,
while other clusters with even sites have only one M point in the BZ. As a consequence of that,
the square sublattice magnetization for the stripe phase on the 36-site torus is overestimated compared to other
system sizes (such as 18, 24, 30), as can be seen in Fig.~\ref{fig:MagneticOrders} of the main text.
One should also note that the 24-site cluster we use here is different from those in Ref.~\cite{Kawamura2015}.

For the tori smaller than or equal to 30 sites, we use exact diagonalization to do the calculations,
while for the 36- and 48-site clusters we use SU(2) DMRG by keeping as many as $8000$ U(1)-equivalent
states to do the calculations. The truncation errors are less than $5\times 10^{-5}$ in all calculations.

\begin{figure}
  \centering
  \includegraphics[width=0.48\textwidth]{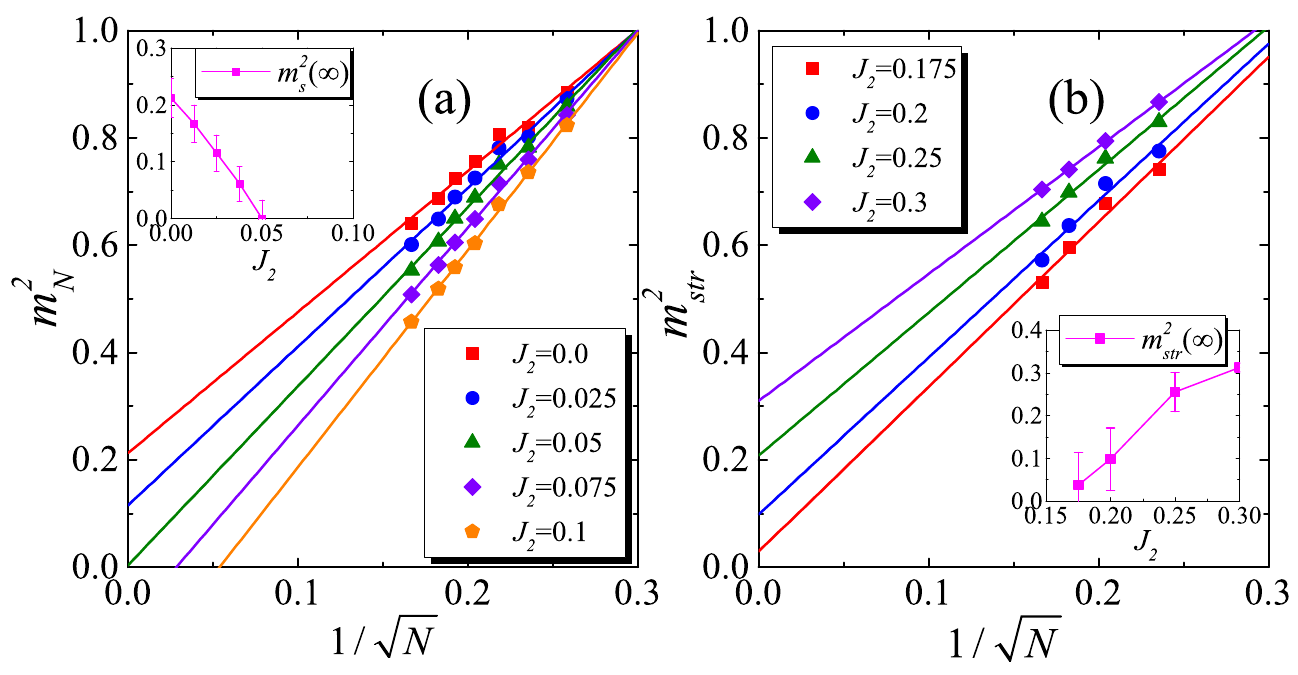}
  \caption{Linear finite-size scaling of square magnetization of (a) $120^{0}$ AF order and (b) stripe
  AF order vs $1/\sqrt{N}$ at various next-nearest-neighbor interactions $J_{2}$. The insets are
  the extrapolated values in the thermodynamic limit.}
  \label{fig:AppMagnetic}
\end{figure}

\section{$J_{1}-J_{2}$ Triangular Heisenberg model}
\label{App:J1J2Model}

\begin{figure}
  \centering
  \includegraphics[width=0.45\textwidth]{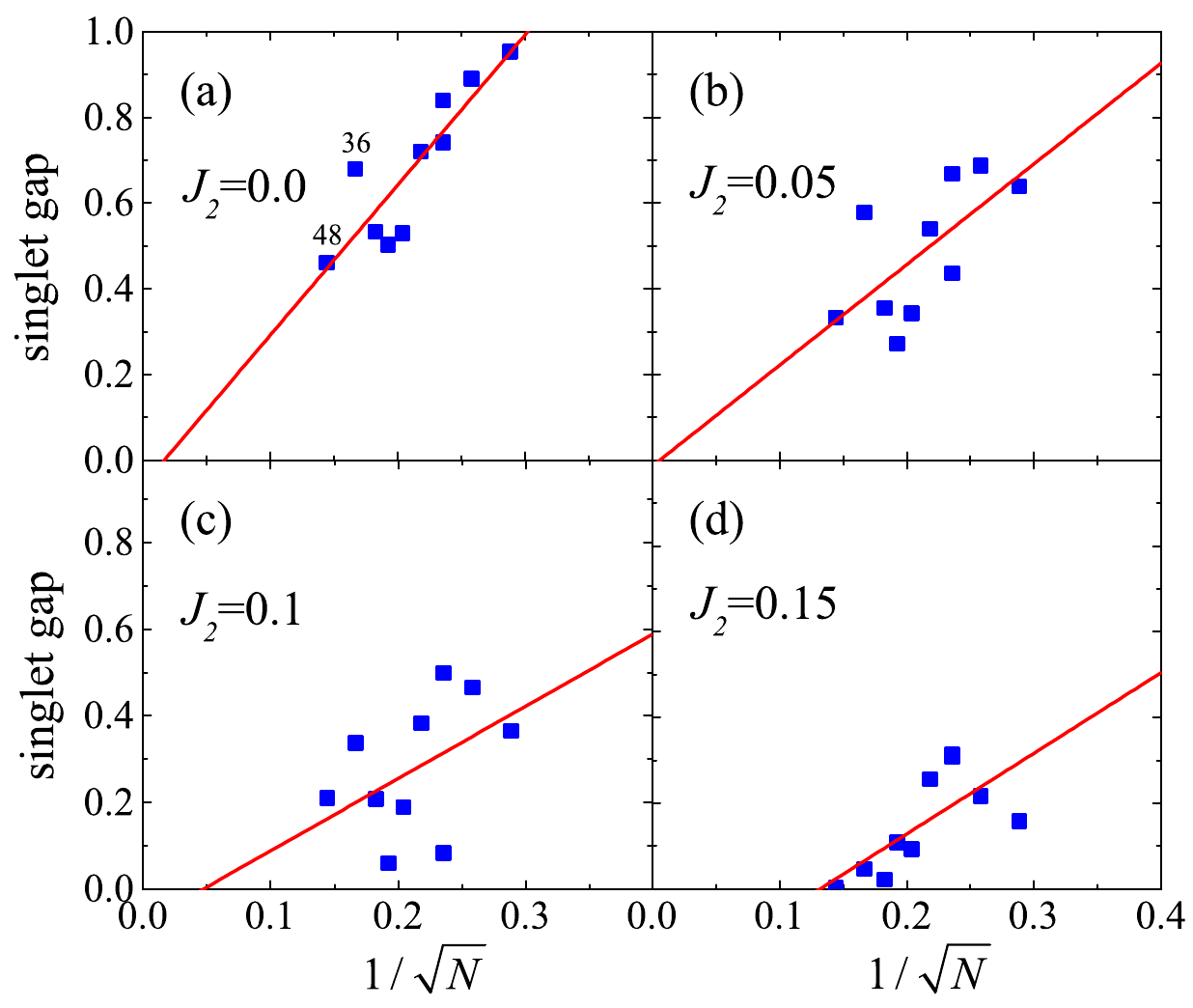}
  \caption{Linear extrapolation of singlet gaps with $1/\sqrt{N}$ at various $J_{2}$. The solid lines
  are least-squares fitting lines. The singlet gaps all seem to be zero in the thermodynamic limit.}
  \label{fig:AppSinglet}
\end{figure}

\begin{figure}
  \centering
  \includegraphics[width=0.45\textwidth]{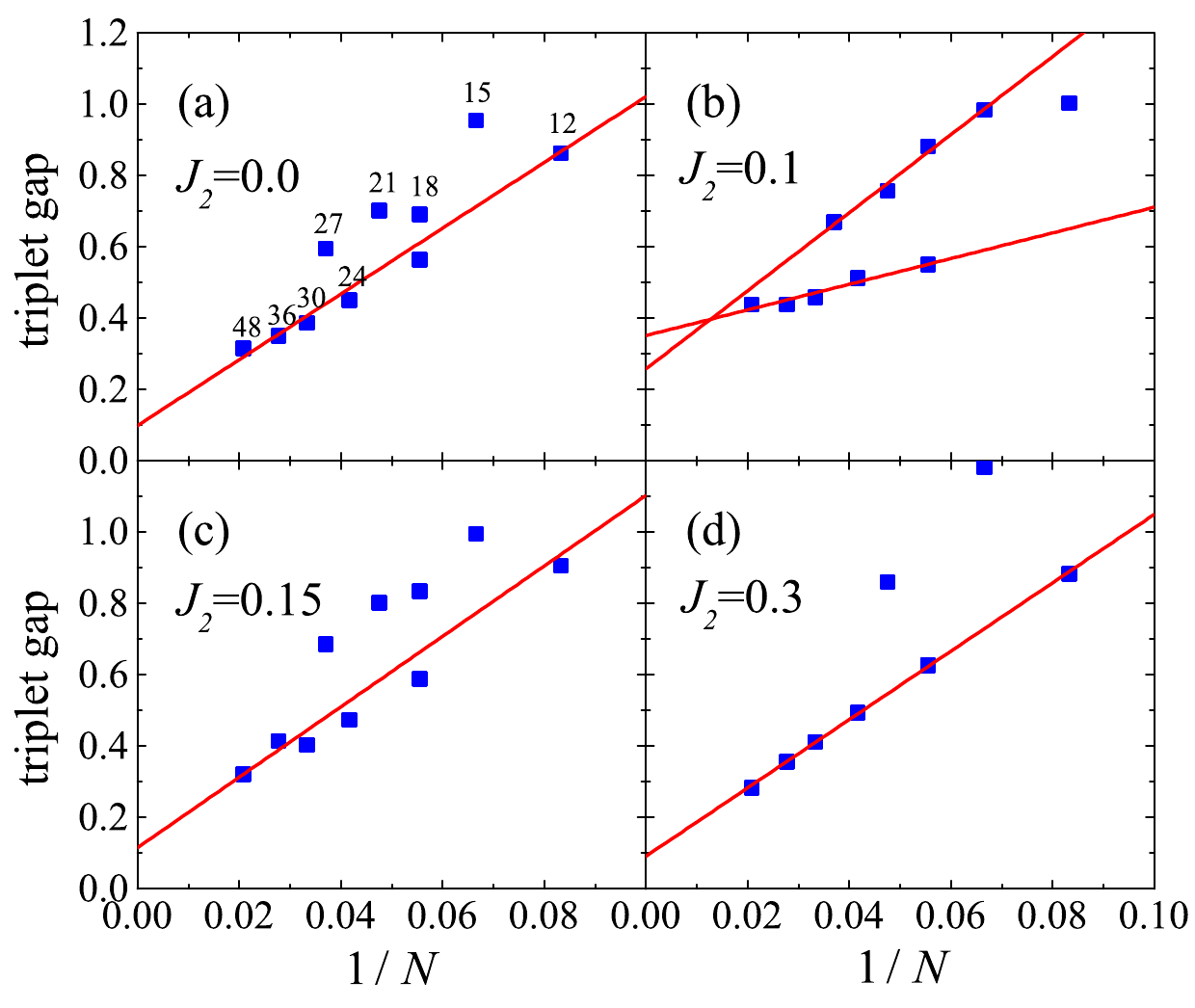}
  \caption{Triplet gaps scale with $1/N$ at various $J_{2}$. The solid lines in (a) and (c) are least-squares fitting
  lines using even-size tori. At $J_2=0.1$, we use two groups of data to do the fitting and ignore the small 12-site torus. Two 18-site tori are used here, one is illustrated in Fig.~\ref{fig:AppClusters} and the other is a rhombic cluster expanded by $6 \times 3$ primitive cells. The 18-site rhombic cluster has smaller triplet gaps than the nonrhombic cluster.}
  \label{fig:AppTriplet}
\end{figure}

We have used finite-size tori to study the non-randomness $J_{1}-J_{2}$ Heisenberg model on
the triangular lattice. Using linear extrapolation of magnetic order parameters, we determined the nonmagnetic
region which is about $0.05(1) < J_{2} < 0.16(2)$. This phase region is similar to the previous
DMRG results~\cite{DNSheng2015, SRWhite2015} and is larger than the varitonal Monte Carlo (VMC) results~\cite{Imada2014}.

Both the $120^{0}$ AF phase and stripe antiferromagnetic phase spontaneously break the spin SU(2) continuous
symmetry in the thermodynamic limit. According to the Nambu-Goldstone theorem, the system in these
magnetic phase regions has gapless excitations. In finite-size systems, a characteristic and
systematic structure of the continuous symmetry breaking is the Anderson tower of states (TOS)
in the energy spectrum. The TOS energy levels scales with $1/N$ to the ground state, while the
low energy magnon excitations scale with $1/\sqrt{N}$ (or $1/L$, $L$ is the linear system size).
Based on that knowledge, we scale the singlet gap with $1/\sqrt{N}$ and triplet gap with $1/N$ where $N$ is the number of lattice sites.

In the SU(2) symmetry breaking phases, the singlet and triplet gaps should go to zero in the
thermodynamic limit in the magnetic regions. From our finite-size calculations, though some
data has large variance, we still can see the gapless tendency in Fig.~\ref{fig:AppSinglet}
(a-b) and Figs.~\ref{fig:AppTriplet} (a) and \ref{fig:AppTriplet}(d). Unfortunately, the system size is still not large enough to
unbiasly extrapolate the triplet gap to zero in the finite-size scaling.
For the nonmagnetic phase [Figs.~\ref{fig:AppSinglet} (c), \ref{fig:AppSinglet}(d) and Figs.~\ref{fig:AppTriplet} (b), \ref{fig:AppTriplet}(c)], it is even harder to draw a
conclusion whether it is gapless or not using the finite-size clusters and linear extrapolation.

\section{Dimer correlation}
\label{App:DimerCorr}

\begin{figure}[htbp]
  \centering
  \includegraphics[width=0.5\textwidth]{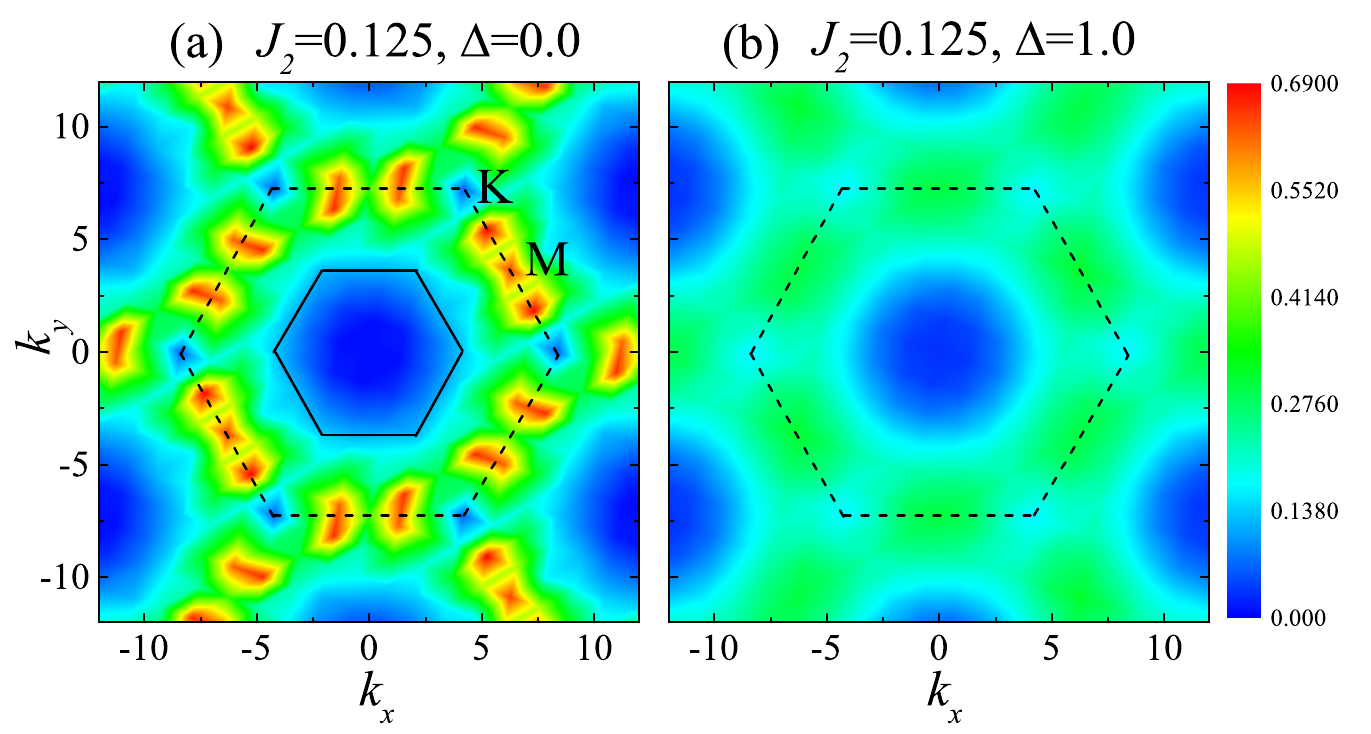}
  \caption{Contour plot of dimer correlation in momentum space at $J_{1}=0.125, \Delta=0.0$ and $J_{1}=0.125, \Delta=1.0$.}
  \label{fig:AppDimET3D}
\end{figure}

In this sector, we show some dimer-dimer correlation function in momentum space. In order to
see the possible off-diagonal valence bond solid pattern, we take every bond as a new lattice
site which is sitting in the middle of each bond. These new sites form a kagome lattice (1/4-depleted
triangular lattice, dashed lines in Fig.~\ref{fig:AppClusters}) or with $3N$ lattice sites, $N$ is
the number of sites in the original triangular lattice. Then we take the Fourier transform from real
space to momentum space using Eq.~\eqref{eq:DimerBi}. Here, we show the contour plot of dimer
correlation in momentum space using the 24-site cluster, which is shown in Fig.~\ref{fig:AppDimET3D}.
We take the maximum $D(X)$ to do the structure factor scaling. $X$ is the momentum site where
$D(\mathbf{q})$ takes its maximum. And it is the same or close to the middle point in between
K and M points [see Fig.~\ref{fig:AppDimET3D} (a)]
depending on the geometry of the finite-size clusters. There is no pattern of long-range valence-bond-solid (VBS)
order in our numerical study (see Fig.~\ref{fig:ChiETDimET} in the main text). In Fig.~\ref{fig:AppDimET3D}
(a), the solid hexagon is the Brillouin zone edge of the original triangular lattice with $N$ sites,
while the dashed hexagon is the ``Brillouin zone" edge of the new depleted triangular lattice with $3N$ sites.

\section{Histogram of spin correlations under different bond randomness strength}
\label{App:Histogram}

\begin{figure}[htbp]
  \centering
  \includegraphics[width=0.5\textwidth]{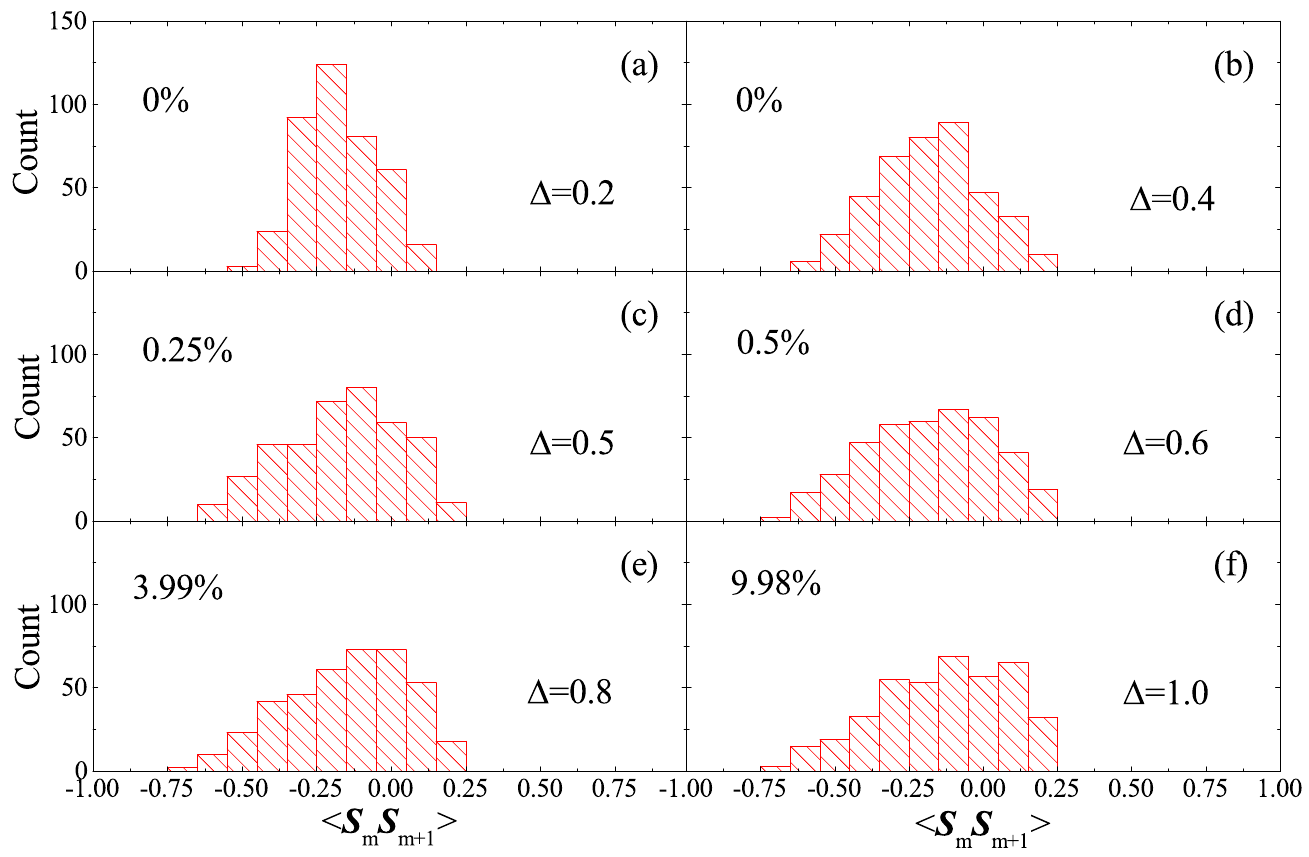}
  \caption{Histograms of nearest-neighbor spin correlation with different bond randomness strengths
  $\Delta$. The finite-size system we take is 24-site torus with 400 independent disorder configurations,
  and the next-nearest-neighbor exchange interaction $J_2$ is set to be 0.125$J_1$. The percentages
  shown in the boxes mean the proportions of triplet ground state under 400 disorder configurations.}
  \label{fig:AppHisto}
\end{figure}

\begin{figure}[htbp]
  \centering
  \includegraphics[width=0.5\textwidth]{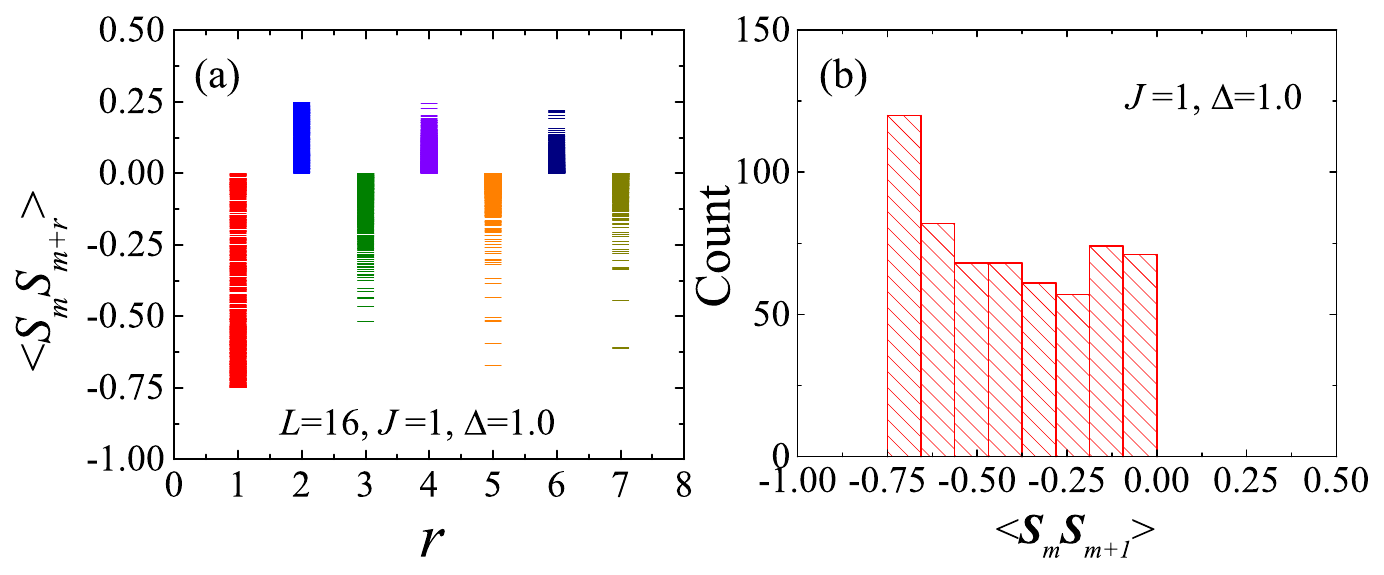}
  \caption{(a) Spin-spin correlations (with distributions) at different distances on the $L=16$ Heisenberg chain with bond randomness $\Delta=1.0$. The logarithmic corrections to the power-law decaying correlations have been found in recent quantum Monte Carlo simulations~\cite{shu2016}. (b) The histogram of nearest-neighbor spin-spin correlation obtained from 600 independent random samples. Two nearest-neighbor spins have a large probability to form a singlet with the correlations trending to $-\frac{3}{4}J$.}
  \label{fig:App1DRS}
\end{figure}

Here, we want to show how the distribution of nearest-neighbor (n.n.) spin correlation changes with the
bond randomness strength. As the bond randomness strength increases, the distribution of n.n.
spin correlation becomes broaden and extends to $-\frac{3}{4}J$ and $\frac{1}{4}J$. Also, the distribution changes
from a gaussian-like shape to a asymmetric one. However, it is a rare event to be a (approximated)
singlet between two nearest-neighbor sites. It is an striking difference between 1D random singlet phase
and the SLL phase. In 1D RS phase, the n.n. spin correlation has large probability to be $-\frac{3}{4}J$ in Fig.~\ref{fig:App1DRS}.

\bibliography{tri}

\end{document}